\documentclass[11pt,a4paper]{article}
\pdfoutput=1
\usepackage{jheppub}

\usepackage{multirow, graphicx,amssymb,url,mathrsfs,amsmath}
\usepackage{wrapfig,boxedminipage,setspace,subfigure,epsfig}
\usepackage{amsxtra,amstext,latexsym,dsfont,amsfonts}
\usepackage{color}
\usepackage[dvipsnames]{xcolor}
\usepackage{float}
\usepackage{slashed,comment}
\usepackage{contour}

\usepackage{tikz}
\usetikzlibrary{calc,patterns,angles,quotes}
\usetikzlibrary{decorations.pathreplacing,decorations.markings,snakes}

\usepackage{tabularx, array}
\newcolumntype{L}[1]{>{\raggedright\arraybackslash}p{#1}}
\newcolumntype{C}[1]{>{\centering\arraybackslash}p{#1}}
\newcolumntype{R}[1]{>{\raggedleft\arraybackslash}p{#1}}

\newcommand{\dd}{\mathrm{d}}

\newcommand{\bra}[1]{\mbox{$\langle #1 |$}}
\newcommand{\ket}[1]{\mbox{$| #1 \rangle$}}

\newcommand{\Tr}{{\rm Tr}}

\usepackage{xparse}

\title{Cosmological pole-skipping, shock waves and quantum chaotic dynamics of de Sitter horizons}

\author[a]{Yongjun Ahn,}
\author[b,c]{Sa\v{s}o Grozdanov,}
\author[d,e,f]{Hyun-Sik Jeong,}
\author[d]{Juan F. Pedraza}

\emailAdd{yongjunahn@sjtu.edu.cn}
\emailAdd{saso.grozdanov@ed.ac.uk}
\emailAdd{hyunsik.jeong@apctp.org}
\emailAdd{j.pedraza@csic.es}

\preprint{\texttt{IFT-UAM/CSIC-25-88, APCTP Pre2025 - 019}}

\affiliation[a]{Wilczek Quantum Center, School of Physics and Astronomy, Shanghai Jiao Tong University, Shanghai 200240, China}
\affiliation[b]{Higgs Centre for Theoretical Physics, University of Edinburgh, Edinburgh, EH8 9YL, Scotland}
\affiliation[c]{Faculty of Mathematics and Physics, University of Ljubljana, Jadranska ulica 19, SI-1000 Ljubljana, Slovenia}
\affiliation[d]{Instituto de F\'isica Te\'orica UAM/CSIC, Calle Nicol\'as Cabrera 13-15, 28049 Madrid, Spain}
\affiliation[e]{Asia Pacific Center for Theoretical Physics, Pohang 37673, Korea}
\affiliation[f]{Department of Physics, Pohang University of Science and Technology, Pohang 37673, Korea}

\abstract{We present a systematic analysis of pole-skipping for scalar, Maxwell, and gravitational waves in cosmological spacetimes. Specifically, working in empty de Sitter space and in Schwarzschild–de Sitter black hole geometries, we locate the tower of pole-skipping points of such fields and show that they impose nontrivial constraints on the corresponding bulk two-point functions. Focusing on the gravitational sound channel, we then extract the Lyapunov exponent and butterfly velocities that characterize hypothetical dual many-body quantum chaos at each horizon. These chaotic data precisely match the outcome of a gravitational shock wave calculation, confirming that the relevant pole-skipping points encode high-energy scattering of horizon quanta. Interestingly, the butterfly velocities can become superluminal or imaginary, with the latter signaling a spatially modulated propagation of chaos. Assuming that a holographic dual exists, we translate our results into field theory language and propose that the dual theory can be divided into two entangled sectors that capture the black hole and cosmological horizon degrees of freedom. Our results suggest that the black hole sector becomes increasingly nonlocal as the black hole shrinks and that the cosmological horizon sector exhibits behavior compatible with violations of Hermiticity. Finally, we outline simple microscopic toy models---built from long-range and non-Hermitian deformations of the Double Scaled Sachdev–Ye–Kitaev (DSSYK)-type chains---that realize these features, providing a concrete arena for future exploration.
}

\begin{document}
\maketitle

%
\section{Introduction}
The holographic principle proposes that the physics of certain gravitational systems can be encoded in properties of their boundaries~\cite{tHooft:1993gx,Susskind:1994vu}. This concept is strongly supported by developments in quantum gravity and information theory, beginning with the Bekenstein-Hawking formula, which relates black hole entropy to the area of the event horizon~\cite{Bekenstein:1972tm,Hawking:1975vcx}. A concrete realization of this principle is provided by the AdS/CFT correspondence~\cite{Maldacena:1997re}, which establishes a duality between quantum gravity in Anti-de Sitter (AdS) space and a conformal field theory (CFT) on its boundary.

\paragraph{De Sitter space and its holographic nature.}
Despite numerous successes of the AdS/CFT correspondence, extending holography to spacetimes with a positive cosmological constant and recovering `simple' workable models remains an open challenge (for an early proposal of dS/CFT correspondence, see Ref.~\cite{Strominger:2001pn}). Presumably, this is due to the fact that constructing convincing top-down string theoretic models with de Sitter space has proven to be extremely difficult. Nevertheless, successful efforts would likely have significant implications for our understanding of cosmology. This is because de Sitter space plays a prominent role in modern astrophysics and cosmology~\cite{SWCosmology}. Observations of the accelerating expansion of our universe (e.g., the astrophysical observations of supernovae \cite{SupernovaSearchTeam:1998fmf,SupernovaCosmologyProject:1998vns}) strongly suggest that our spacetime is approximately described by a homogeneous and isotropic metric with a positive cosmological constant, rendering it close to the dS geometry (see e.g.~Refs.~\cite{Spradlin:2001pw,Anninos:2012qw,Galante:2023uyf} for reviews of dS geometry). Moreover, according to the standard understanding of the $\Lambda$CDM cosmology, we are `presently' entering into the era of the cosmological-constant-dominated (or the dark-energy dominated) universe, which will asymptote to pure de Sitter spacetime in the future.   

A defining characteristic of dS space, in contrast to asymptotically AdS spacetimes, is the existence of a `cosmological horizon' that limits the observable (causal) region for any given inertial observer. Similarly to the event horizon, the cosmological horizon also has both a temperature and an entropy \cite{Gibbons:1977mu}:
\begin{align}\label{}
    T_{\text{dS}} = \frac{1}{2\pi L}  ,\quad  S_{\text{dS}} = \frac{A}{4 G_N}  = \frac{\Omega_{d-1} L^{d-1}}{4 G_N} ,
\end{align}
where $L$ is the dS radius and $\Omega_{d-1}$ the volume of the unit $(d-1)$ sphere. In analogy with black holes, these quantities naturally raise the possibility that dS space also exhibits a `holographic nature'. However, cosmological horizons are inherently observer-dependent, making any holographic description more subtle. While black holes are localized and admit observers at asymptotic infinity, dS observers have limited causal reach and experience distinct horizons. The central role of the observer in dS physics has been emphasized in recent works~\cite{Chandrasekaran:2022cip,Witten:2023qsv}; for a summary of associated quantum issues involving the cosmological horizon, see Ref.~\cite{Bousso:2002fq}.

In statistical mechanics, entropy reflects the number of microstates consistent with the macroscopic properties of a system. Applied to dS space, the central conjecture---often referred to as the dS central dogma---suggests that the cosmological horizon encodes a unitary quantum system with approximately $\exp (A/4G_N)$ states~\cite{Susskind:2021esx,Shaghoulian:2021cef,Shaghoulian:2022fop,Bousso:1999dw,Banks:2001yp,Banks:2002wr,Dyson:2002nt,Banks:2006rx,Banks:2018ypk,Susskind:2021omt}. This implies that any such quantum theory dual to dS space should have a finite number of degrees of freedom and be described by a maximally mixed state~\cite{Chandrasekaran:2022cip,Dong:2018cuv,Lin:2022nss}. A widely explored framework in this context is the static patch holography~\cite{Baiguera:2023tpt}, in which the holographic dual theory is localized on a constant radial slice just inside the cosmological horizon, i.e., within the causal region accessible to a static observer in dS space.\footnote{In three bulk dimensions, a more concrete realization of the dual theory has been proposed using a $T\bar{T}$ deformation~\cite{Lewkowycz:2019xse}. This setup allows for an explicit counting of the microstates associated with the cosmological horizon~\cite{Coleman:2021nor}, consistent with earlier results~\cite{Anninos:2020hfj}. Extensions to higher dimensions have also been investigated, e.g., in Ref.~\cite{Silverstein:2022dfj}.}

Beyond static patch holography, several alternative approaches to `dS holography' have been proposed. As already noted above, one of the earliest such attempts is the dS/CFT correspondence~\cite{Strominger:2001pn}, which conjectures a duality between quantum gravity in $d+1$-dimensional asymptotically dS space and a Euclidean $d$-dimensional CFT living at timelike infinity. Another strategy involves embedding dS space into a spacetime with a timelike boundary, thereby enabling the application of AdS/CFT tools. For such constructions see Refs.~\cite{Freivogel:2005qh,Lowe:2010np,Fischetti:2014uxa,Anninos:2017hhn,Anninos:2018svg,Mirbabayi:2020grb,Anninos:2020cwo}. Nevertheless, a complete holographic formulation of dS space remains an open challenge. A precise identification of the dual quantum theory---hypothetically defined at the boundary of dS space---has not yet been achieved. Moreover, it is unclear how the well-established AdS/CFT correspondence and its dictionary of mappings can be adapted or extended to the dS context.

Without resolving any of these detailed (microscopic) problems of de Sitter holography, in this work, our aim is to investigate potential `universal' structures that constrain the linearized bulk gravitational perturbations in dS space. In particular, we investigate the pole-skipping phenomenon \cite{Grozdanov:2017ajz,Blake:2017ris,Blake:2018leo}, its connections to quantum chaotic behavior of a hypothetical dual theory, and compare its signatures at the black hole event horizon and the cosmological horizon in asyptotically de Sitter space. 

\paragraph{Horizon constraints in dS space: cosmological pole-skipping.} In the context of AdS spacetime, it is well known that the presence of black hole horizons imposes nontrivial constraints on physical observables, e.g., on bulk two-point functions and the corresponding dual Green’s functions of various operators. A clear example of this is the above-mentioned phenomenon known as pole-skipping \cite{Grozdanov:2017ajz,Blake:2017ris,Blake:2018leo}, in which specific points in the complex frequency-momentum plane correspond to non-unique solutions of the bulk field equations. Notably, at these special pole-skipping points, the boundary two-point function takes the value of `$0/0$' meaning that its (regular) value depends on the limit of approach to the point. It has been firmly established that for `maximally chaotic' theories with dual classical gravity descriptions, this phenomenon is related to quantum chaos. The so-called `chaotic' pole-skipping point allows for a computation of the butterfly velocity otherwise extracted from the out-of-time-ordered correlators (OTOC).\footnote{For a recent proof of this statement, see Ref.~\cite{Chua:2025vig}.}

Beyond pole-skipping associated with holographic maximal quantum chaos at $\omega = i \lambda_L = 2\pi T i$, where $\lambda_L = 2 \pi T$ is the maximal Lyapunov exponent, the existence of an infinite set of pole-skipping points at $\omega = -2 \pi T i n$, with integer $n \geq 0$, has since been recognized as a generic feature of thermal correlation functions \cite{Grozdanov:2019uhi,Blake:2019otz} (for an earlier work identifying such points, see Ref.~\cite{Amado:2008ji}). The phenomenon has also been explored in `generic' strongly chaotic systems~\cite{Blake:2017ris,Blake:2021wqj,Jeong:2021zhz}, beyond the maximally chaotic regime of the SYK chain~\cite{Choi:2020tdj}, in CFTs~\cite{Haehl:2018izb,Jensen:2019cmr,Das:2019tga,Haehl:2019eae,Ramirez:2020qer}, and also in quantum mechanical systems~\cite{Natsuume:2021fhn}. The breadth of such examples indicates its broad applicability. A number of works have also focused on establishing the mathematical structures and universality of pole-skipping points, for instance, in Refs.~\cite{Blake:2019otz,Natsuume:2019xcy,Natsuume:2019sfp,Natsuume:2019vcv,Ceplak:2019ymw,Ahn:2019rnq,Ahn:2020bks,Abbasi:2020ykq,Liu:2020yaf,Ahn:2020baf,Natsuume:2020snz,Sil:2020jhr,Ceplak:2021efc,Blake:2021hjj,Ageev:2021xjk,Wang:2022xoc,Wang:2022mcq,Amano:2022mlu,Grozdanov:2023txs,Jeong:2023rck,Grozdanov:2023tag,Jeong:2023ynk,Abbasi:2023myj,Pan:2024azf,Ahn:2024gjh,Grozdanov:2024wgo,Grozdanov:2025ner,Lu:2025jgk,Lu:2025pal,Lilani:2025wnd}.

Similar techniques can also be applied to investigate the linearized perturbations of fields in asymptotically de Sitter spacetimes. For an example of such an analysis in four-dimensional gravity, see Ref.~\cite{Grozdanov:2023txs}. In this paper, we investigate the phenomenon of pole-skipping in asymptotically de Sitter black holes more broadly and discuss the constraints that dS horizons impose on perturbations of gravity, the Maxwell field and scalars. Of primary focus will be the Schwarzschild-de Sitter (SdS) black hole. We systematically show that, as in AdS, at specific frequencies and momenta, the bulk equations admit non-unique solutions at the black hole event horizon and the cosmological horizon, signaling the existence of `cosmological' pole-skipping in associated Green’s functions of a hypothetical dual theory. Beyond identifying the locations of pole-skipping points, we also aim to explore the physical interpretation of the special `chaotic' pole-skipping point that in AdS holography exists at $\omega = i \lambda_L$ and allows to compute the associated butterfly velocity. Of particular interest will be such behavior at the cosmological horizon, which we compare to the analysis of an OTOC-like shockwave analysis. Furthermore, in discussing potential dual physical interpretations of cosmological pole-skipping, we propose a possible microscopic interpretations of our results. This is done by suggesting connections to models such as the double-scaled Sachdev-Ye-Kitaev (DSSYK) chain and non-Hermitian deformations. 

In essence, by systematically analyzing how pole-skipping manifests in dS geometries and comparing it to the better-understood AdS case, we aim to enhance our understanding of the interplay between horizon constraints, quantum chaos, and field theoretic observables in spacetimes with a positive cosmological constant: an area of growing relevance for future astrophysical and cosmological studies.

The paper is organized as follows. In Section~\ref{sec2}, we review the key properties of asymptotically dS geometries, focusing on empty de Sitter and Schwarzschild-de Sitter black holes. Section~\ref{sec3} details the near-horizon method for computing the pole-skipping points of scalar field, the Maxwell field, and gravitational perturbations. In Section~\ref{sec4}, we interpret these results in terms of another bulk diagnostic of chaos: the shockwave geometry. Of particular interest will then be the interpretation of the hypothetical `butterfly velocities' for which we outline possible microscopic interpretations, including hints from the DSSYK chain and the role of non-Hermitian deformations. We conclude in Section~\ref{sec5} with a summary of our results and future outlook.

%
\section{Preliminaries: the background spacetime}\label{sec2}

We begin with a review of the key properties of asymptotically de Sitter (dS) spacetimes, covering both the empty dS case and the Schwarzschild-de Sitter (SdS) black holes.

\subsection{Asymptotically de Sitter spacetimes}
We consider the ($d+1$)-dimensional Einstein-Hilbert action ${S}$ with positive a cosmological constant:
\begin{align}\label{Action:EH}
{S} = \frac{1}{16\pi G_N}\int_\mathcal{M} \dd^{d+1}x \left(R - 2 \Lambda\right) , \quad  \Lambda=\frac{d(d-1)}{2L^2} ,
\end{align}
where $G_N$ is the Newton's gravitational constant, $\Lambda$ the cosmological constant, and $L$ the dS space length scale. For the metric tensor, we use the ansatz
\begin{align}\label{ORIMET}
    \dd s^2 = g_{\mu\nu} \dd x^{\mu} \dd x^{\nu} = g_{ab}\dd x^a \dd x^b + r^2  \gamma_{ij}\dd x^i \dd x^j ,
\end{align}
where the $(d+1)$-dimensional manifold is decomposed as $\mathcal{M}=\mathcal{N}^2 \times \mathcal{K}^{d-1}$. Here, $\mathcal{N}^2$ is a two-dimensional Lorentzian spacetime with metric $g_{ab}$, while $\mathcal{K}^{d-1}$ is a $(d-1)$-dimensional maximally symmetric Riemannian space with metric $\gamma_{ij}$. In static coordinates, the metric takes the form
\begin{align}\label{Ansatz:INEF}
\begin{split}
g_{ab}\dd x^a \dd x^{b} = -f(r)\dd t^2 +\frac{1}{f(r)}\dd r^2,\quad
    \gamma_{ij}\dd x^i \dd x^j =\frac{\dd \chi^2}{1 - \chi^2}+ \chi^2 \dd \Omega^2_{d-2} ,
\end{split}
\end{align}
where $\dd \Omega^2_{d-2}$ is the line element of a unit sphere $\mathcal{S}^{d-2}$, defined iteratively as
\begin{align}\label{AGD}
\begin{split}
   \dd \Omega_1^2 = \dd \theta_1^2 , \quad \dd \Omega^2_{j} = d\theta_{j}^2+\sin^2(\theta_j)  \dd \Omega^2_{j-1} ,
\end{split}
\end{align}
with angular coordinates $\theta_{1} \in [0, 2\pi]$ and $\theta_{j} \in [0, \pi]$ for $j \in \{2, \cdots, d-2\}$. 
The blackening factor $f(r)$ is given by~\cite{Spradlin:2001pw}
\begin{align}\label{Sol:f}
    f(r) = 1 - \frac{2M}{r^{d-2}} - \frac{r^2}{L^2} ,
\end{align}
where $M$ is a parameter related to the mass of the solution. The case $M=0$ corresponds to empty dS space, while $M\neq0$ describes the SdS black holes. It is important to note that in asymptotically dS spacetimes, black holes can only have spherical horizons (so that $\mathcal{K}^{d-1}=\mathcal{S}^{d-1}$), which constrains the possible horizon geometries.

The function $f(r)$ in Eq.~\eqref{Sol:f} typically has two roots corresponding to the black hole event horizon $r_\mathrm{bh}$ and the cosmological horizon $r_\mathrm{c}$, with $r_\mathrm{bh} < r_\mathrm{c}$. Each horizon is associated with a temperature given by~\cite{Gibbons:1977mu}\footnote{The associated entropy is $S_{\mathrm{bh}|\mathrm{c}} = {\Omega_{d-1} r^{d-1}_{\mathrm{bh}|\mathrm{c}}}/{4G_{N}}$.}
\begin{align}\label{TSfor}
    T_{\mathrm{bh}|\mathrm{c}} = \bigg|\frac{f'(r)}{4\pi}\bigg|_{r=r_{\mathrm{bh}|\mathrm{c}}} ,
\end{align}
where by the notation $\mathrm{bh}|\mathrm{c}$, we mean either the black hole event horizon or the cosmological horizon. In the following subsections, we further examine the asymptotically dS spacetime for different spacetime dimension and the mass parameter.

\subsection{Empty de Sitter space}
Let us start with empty de Sitter (dS) space in $(d+1)$ dimensions. Setting $M=0$ in Eq.~\eqref{Sol:f}, the blackening factor in any number of dimensions simplifies to
\begin{align}\label{}
    f_{\text{dS}}(r) = 1 - \frac{r^2}{L^2} .
\end{align}

\begin{figure}[]
  \centering
     {\includegraphics[width=6.4cm]{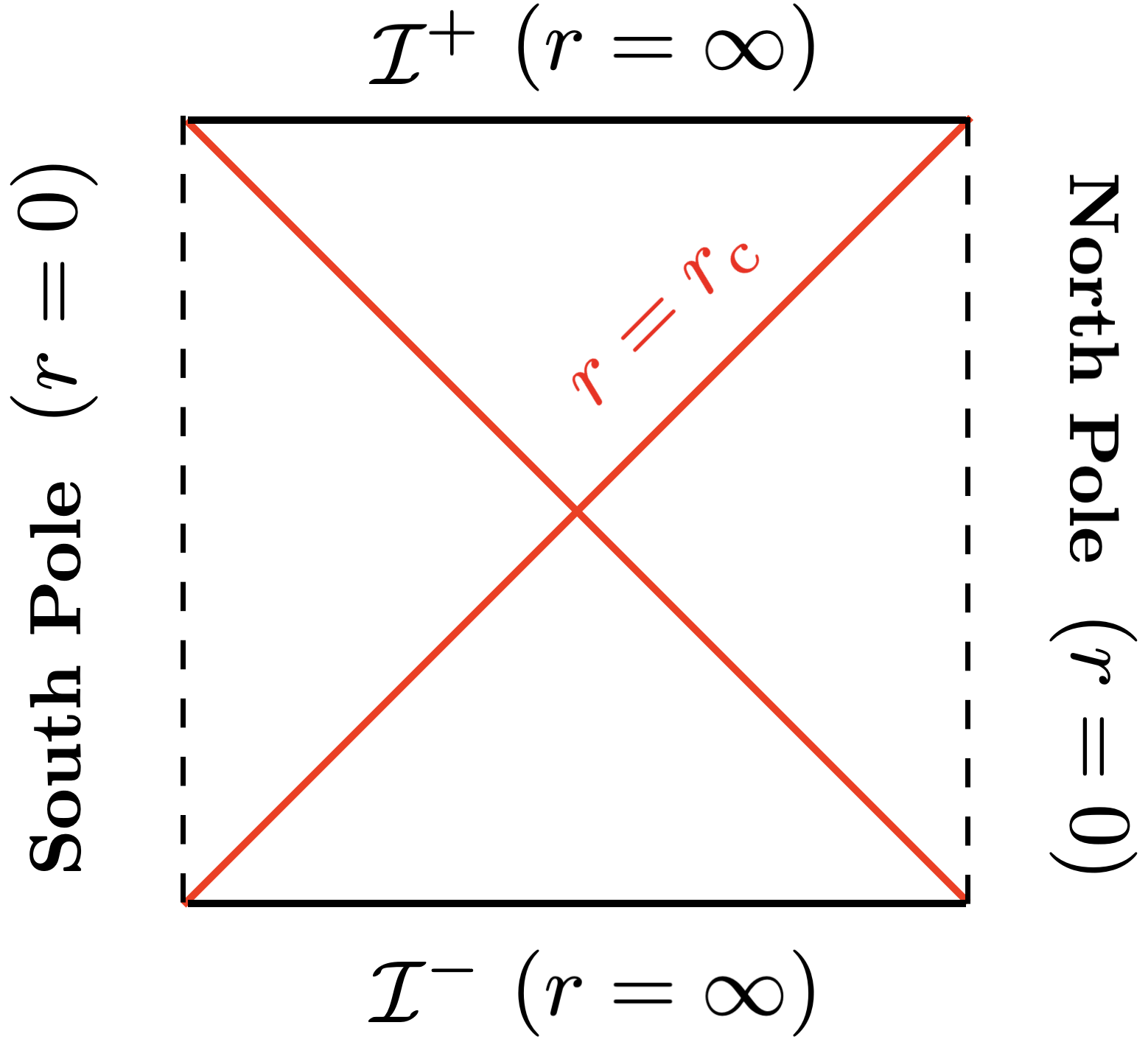} \label{}}
 \caption{Penrose diagram of empty dS space. The red line denotes the cosmological horizon at $r=r_\mathrm{c}=L$, the solid black lines represent future and past timelike infinities $\mathcal{I}^{\pm}$ at $r=\infty$, and the dashed black lines correspond to the worldlines of an observer at the south pole (left) and the north pole (right), located at $r=0$.}\label{PENROSEFIG1}
\end{figure}

The spacetime topology of empty dS space is $\mathbb{R} \times \mathcal{S}^{d}$, with its Penrose diagram shown in Figure~\ref{PENROSEFIG1}. The dashed black lines on either side of the diagram at $r=0$ represent the worldlines of inertial observers at the spatial poles of $\mathcal{S}^{d}$.
The coordinate system in Eq.~\eqref{Ansatz:INEF} then describes the static patch of dS space, which is the region causally accessible to an observer situated at either pole. Due to the presence of a positive cosmological constant, this region is bounded by a cosmological horizon at $r = r_\mathrm{c} = L$.

The dS background is associated with a finite temperature and entropy due to the presence of the cosmological horizon. Using Eq.~\eqref{TSfor}, its Hawking temperature is
\begin{align}\label{dSTEM}
    T_{\text{dS}} = \frac{1}{2\pi L}  .
\end{align}

\subsection{Schwarzschild-de Sitter black hole}
The Schwarzschild-de Sitter (SdS) spacetime, given by Eq.~\eqref{Sol:f} with $M\neq0$, is a neutral, static, and spherically symmetric solution of Einstein's equations in an asymptotically dS$_{d+1}$ background. It is sometimes referred to as the Kottler metric~\cite{Kottler:1918cxc}. In this manuscript, we focus on the cases with $d\geq2$, where the number of dimensions significantly influences the conformal structure of the spacetime. Notably, for $d=2$, the geometry features \textit{only} a cosmological horizon ($r_\mathrm{c}$), whereas for $d\geq3$, a black hole event horizon ($r_\mathrm{bh}$) generally exists as well. Due to this distinction, we consider these two cases separately.

\paragraph{The $(2+1)$-dimensional case.}
For $d=2$, the metric \eqref{Ansatz:INEF} can be written as
\begin{align}
    \dd s^2 = -f(r)\dd t^2 +\frac{1}{f(r)}\dd r^2 + r^2 \dd \theta^2 , \quad \chi = \cos \theta ,
\end{align}
where the blackening factor \eqref{Sol:f} is
\begin{align}\label{}
    f(r) = 1 - 2M - \frac{r^2}{L^2} .
\end{align}
Here, $M/4G_N$ corresponds to the energy of the solution~\cite{Spradlin:2001pw,Deser:1983nh}. The cosmological horizon is located at 
\begin{align}\label{}
    r_\mathrm{c} = L  \sqrt{1-2M}  ,
\end{align}
whereas, as noted above, no black hole event horizon exists. This result follows from the fact that in $d=2$, the SdS$_{3}$ spacetime is locally identical to empty dS$_3$. Explicitly, by introducing the rescaled coordinates
\begin{align}\label{}
    \tilde{t} = t  \sqrt{1-2M}  ,\quad \tilde{r} = \frac{r}{\sqrt{1-2M}} ,\quad \tilde{\theta} = \theta  \sqrt{1-2M} ,
\end{align}
it becomes evident that the SdS$_3$ metric reduces to empty dS$_3$,
\begin{align}
    \dd s^2 = -\left(1-\frac{\tilde{r}^2}{L^2}\right) \dd \tilde{t}^2 +\frac{1}{\left(1-\frac{\tilde{r}^2}{L^2}\right)}\dd \tilde{r}^2 + \tilde{r}^2 \dd \tilde{\theta}^2  .
\end{align}
with the cosmological horizon at $\tilde{r}_\mathrm{c} = L$. Thus, the conformal structure of the $(2+1)$-dimensional black hole solution is identical to that of empty dS, which we shown in Figure~\ref{PENROSEFIG1}.\footnote{Nevertheless, due to the periodic identification of the new angular coordinate with period $2\pi  \sqrt{1-2M}$, a conical singularity appears at the origin, characterized by a deficit angle of $2\pi \left(1-\sqrt{1-2M}\right)$.} Finally, using Eq.~\eqref{TSfor}, the temperature associated with $r_\mathrm{c}$ is given by
\begin{align}\label{}
    T_{\text{SdS}_3} = \frac{\sqrt{1-2M}}{2\pi L}  .
\end{align}

\begin{figure}[]
  \centering
     {\includegraphics[width=9.4cm]{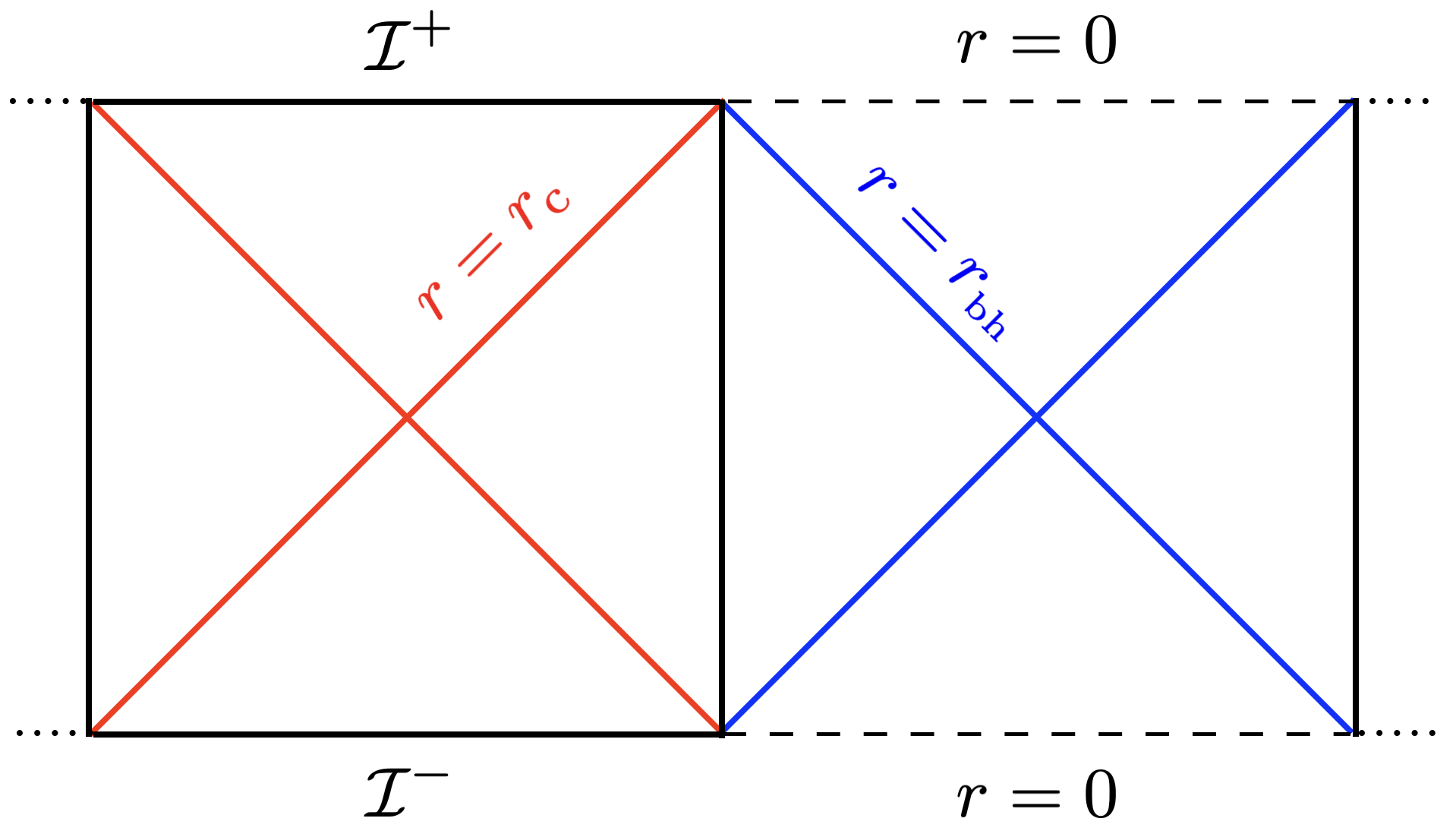} \label{}}
 \caption{Penrose diagram of SdS$_{d+1}$ black holes when $d\geq3$. The red line represents the cosmological horizon ($r=r_\mathrm{c}$), while the blue line denotes the black hole event horizon ($r=r_\mathrm{bh}$).}\label{PENROSEFIG2}
\end{figure}

\paragraph{The higher-dimensional case.}
In dimensions $d\geq3$, and for the mass values in the range of $0<M<M_\text{cr}$, where
\begin{align}\label{CRITICAL}
M_{\text{cr}} \equiv \frac{r_{\text{cr}}^{d-2}}{d}, \quad r_{\text{cr}}\equiv L\sqrt{\frac{d-2}{d}} ,
\end{align}
the spacetime features two distinct horizons: a black hole event horizon at $r_\mathrm{bh}$ and a cosmological horizon at $r_\mathrm{c}$, satisfying $r_\mathrm{bh} < r_\mathrm{c}$.\footnote{Different approaches to defining the mass in this setup can be found in~\cite{Gibbons:1977mu,Ghezelbash:2001vs}.} The corresponding Penrose diagram is now shown in Figure~\ref{PENROSEFIG2}.

The critical value $M_{\text{cr}}$ in \eqref{CRITICAL} corresponds to the scenario in which the blackening factor develops a double root, meaning the two horizons coincide at $r=r_\mathrm{bh} = r_{\text{cr}}$, and
\begin{align}\label{}
f(r_\text{cr}) = 0 , \quad f'(r_\text{cr}) = 0 .
\end{align}
In this limit, the geometry is known as the Nariai solution (see e.g.~Ref.~\cite{Anninos:2012qw}).\footnote{In this limit, the static patch reduces to the geometry of dS$_2\times \mathcal{S}^{d-2}$~\cite{Anninos:2012qw,Maldacena:2019cbz}.}

While a closed-form solution for the horizons is generally unavailable (except in $d=3$, as discussed below), the conditions $f(r_\mathrm{bh})=f(r_\mathrm{c})=0$ allow the mass and the dS curvature scale to be expressed \cite{Morvan:2022ybp} in terms of $r_\mathrm{bh}$ and $r_\mathrm{c}$ as
\begin{align}\label{MLRE}
    M = \frac{r_\mathrm{c}^d r_\mathrm{bh}^{d-2}-r_\mathrm{bh}^d r_\mathrm{c}^{d-2}}{2\left(r_\mathrm{c}^d-r_\mathrm{bh}^d\right)}, \quad L^2=\frac{r_\mathrm{c}^d-r_\mathrm{bh}^d}{r_\mathrm{c}^{d-2} -r_\mathrm{bh}^{d-2}} .
\end{align}
Substituting these expressions into the blackening factor \eqref{Sol:f} yields
\begin{align}\label{Eq:dSSch}
    f(r)=\frac{1}{r_\mathrm{c}^d-r_\mathrm{bh}^d}\left[r_\mathrm{c}^d \left(1-\frac{r^2}{r_\mathrm{c}^2}-\frac{r_\mathrm{bh}^{d-2}}{r^{d-2}}\right)-r_\mathrm{bh}^d \left(1-\frac{r^2}{r_\mathrm{bh}^2}-\frac{r_\mathrm{c}^{d-2}}{r^{d-2}}\right)\right] .
\end{align}
The Hawking temperatures associated with $r_\mathrm{bh}$ and $r_\mathrm{c}$ are given by
\begin{align}\label{SdST}
    T_\mathrm{bh} = d  \frac{r_{\text{cr}}^2-r_\mathrm{bh}^2}{4\pi r_\mathrm{bh} L^2}, \quad T_\mathrm{c}  = d  \frac{r_\mathrm{c}^2-r_{\text{cr}}^2}{4\pi r_\mathrm{c} L^2} .
\end{align}

It is worth noting that for any mass in the range $0<M<M_{\text{cr}}$, the temperatures satisfy $T_\mathrm{bh} > T_\mathrm{c}$, indicating that the system is out of equilibrium.\footnote{In thermal equilibrium, all parts of a system should have the same temperature. Nevertheless, here there is a temperature gradient between the two horizons. This temperature difference leads to a net flow of energy from the hotter black hole horizon to the cooler cosmological horizon. The black hole may emit more radiation than it absorbs, causing it to gradually evaporate~\cite{Nakarachinda:2021aa,Morvan:2022ybp}.} As $M \rightarrow M_{\text{cr}}$, the two temperatures vanish, signaling the Nariai limit, where the system reaches an equilibrium in an extremal configuration~\cite{Anninos:2012qw,Bousso:1997wi}. A separate equilibrium case may occur when $M=0$, corresponding to empty dS space, where $r_\mathrm{c}$ does not evaporate due to thermal balance with the emitted radiation~\cite{Anninos:2012qw}. 

\begin{figure}[]
  \centering
     {\includegraphics[width=6.8cm]{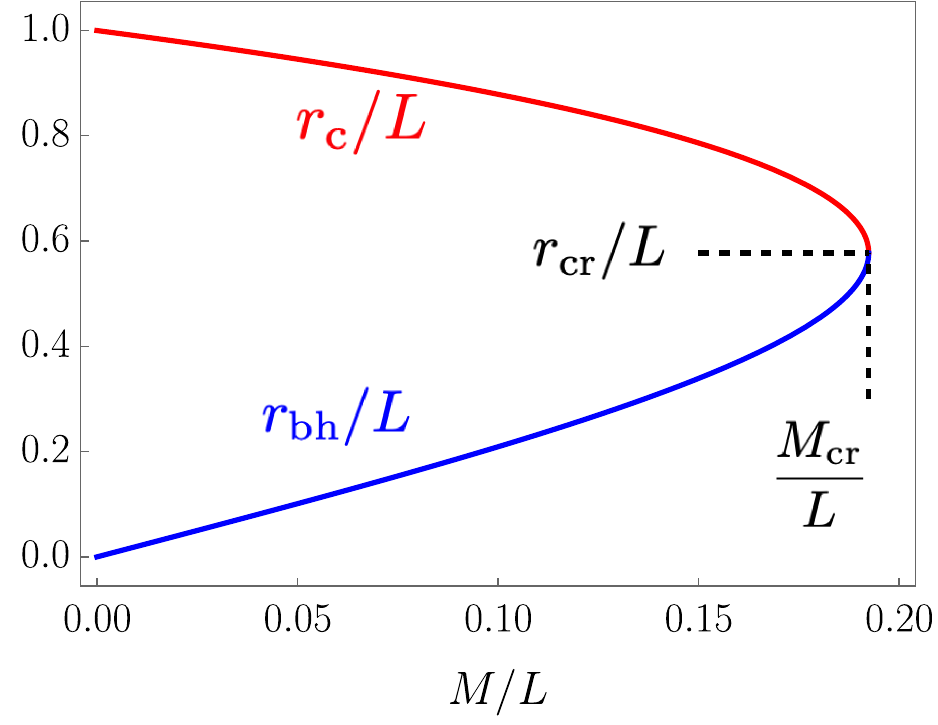} \label{}}
\quad
     {\includegraphics[width=6.8cm]{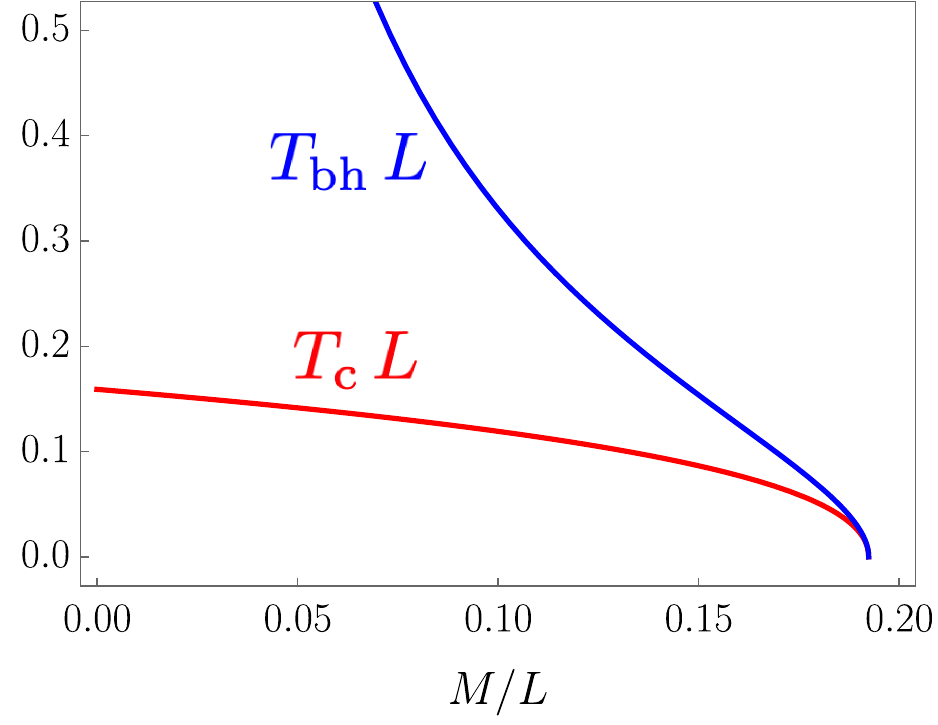} \label{}}
 \caption{\textbf{Left:} The black hole and cosmological horizons in SdS$_{4}$ as functions of $M$ in units of $L$: Eq.~\eqref{rhrcRELATION}. \textbf{Right:} Hawking temperatures of each horizon. When $M=0$, $T_\mathrm{c}$ matches the dS temperature $T_\mathrm{c}  L = 1/2\pi \approx 0.16$: \eqref{dSTEM}.}\label{rhcFIG}
\end{figure}

\paragraph{The special higher-dimensional case in $(3+1)$ dimensions.}
In $d=3$, the blackening factor simplifies to
\begin{align}\label{}
    f(r) = 1 - \frac{2M}{r} - \frac{r^2}{L^2} .
\end{align}
The critical mass \eqref{CRITICAL} takes the form
\begin{align}\label{MCRd3}
\frac{M_{\text{cr}}}{L} = \frac{1}{3\sqrt{3}} \approx  0.19 .
\end{align}
In this case, the mass-horizon relations \eqref{MLRE} can be explicitly inverted \cite{Shankaranarayanan:2003ya,Choudhury:2004ph,Visser:2012wu}:
\begin{align}\label{rhrcRELATION}
    r_\mathrm{bh} = r_{\text{cr}} \left( \cos \eta - \sqrt{3} \sin \eta \right) , \quad 
    r_\mathrm{c} = r_{\text{cr}} \left( \cos \eta + \sqrt{3} \sin \eta \right) , \quad
    \eta \equiv \frac{1}{3}  \arccos \frac{M}{M_{\text{cr}}} ,
\end{align}
with $r_{\text{cr}}=L/\sqrt{3}$. These functions are plotted in the left panel of Figure~\ref{rhcFIG}.
As the black hole mass increases, $r_\mathrm{bh}$ grows while $r_\mathrm{c}$ shrinks, until they meet at $M=M_{\text{cr}}$, forming the Nariai limit. For $M=0$, empty dS space is recovered, with $r_\mathrm{bh}=0$ and $r_\mathrm{c}=L$.

The Hawking temperatures \eqref{SdST} in this case become
\begin{align}\label{}
    T_\mathrm{bh} = \frac{L^2 - 3 r_\mathrm{bh}^2}{4\pi L^2 r_\mathrm{bh}} , \quad T_\mathrm{c}  =  \frac{3 r_\mathrm{c}^2 - L^2}{4\pi L^2 r_\mathrm{c}} ,
\end{align}
which can be written in terms of the mass $M$ by using \eqref{rhrcRELATION}; see the right panel of Figure~\ref{rhcFIG}. For small $M/L$, they asymptotically approach
\begin{align}\label{}
T_\mathrm{bh}  L = \frac{1}{8\pi} \frac{L}{M} + \mathcal{O}\left(\frac{M}{L}\right) , \quad T_\mathrm{c}  L = \frac{1}{2\pi} + \mathcal{O}\left(\frac{M}{L}\right) ,
\end{align}
recovering the empty dS result \eqref{dSTEM} for $T_\mathrm{c}$.
For a small black hole ($r_\mathrm{bh} \rightarrow 0$ or $M \rightarrow 0$), the temperature $T_\mathrm{bh}$ resembles that of a standard flat Schwarzschild black hole, following the relation $T_\mathrm{bh} \approx 1/M$. This inverse dependence on mass implies that as the black hole loses mass through evaporation, its temperature increases, leading to a runaway heating effect.

%
\section{Pole-skipping}\label{sec3}
In this section, we investigate the pole-skipping phenomena of probe scalar, Maxwell, and gravitational field fluctuations in the asymptotically de Sitter background \eqref{ORIMET}. Beyond establishing the generality of pole-skipping for de Sitter geometries, the main purpose of this section is to uncover the `chaotic' pole-skipping points in the upper complex $\omega$ half-plane, which are in AdS spaces related to quantum chaos in the dual QFT.

%
\subsection{Near-horizon methods for pole-skipping: a review}\label{SEC31}

We begin with a quick review of the near-horizon method \cite{Blake:2018leo,Blake:2019otz} to find the pole-skipping points from the local bulk analysis around the event and cosmological horizons. First, it is convenient to introduce the Eddington-Finkelstein (EF) coordinates with $v=t + r_*$, $u=t - r_*$ and the tortoise coordinate is $\dd r_* = {\dd r}/{f(r)}$. We will impose the ingoing bounary conditions both at $r_\mathrm{bh}$ and $r_\mathrm{c}$. The two-dimensional Lorentzian part of the metric $\mathcal{N}^2$ (cf.~Eq.~\eqref{ORIMET}) then takes the form
\begin{align}\label{ansatz:inEF}
    g_{ab}\dd x^a \dd x^b = -f(r)\dd v_{\pm}^2\pm 2 \dd v_{\pm} \dd r ,
\end{align}
where $v_{+} \equiv v$ and $v_{-} \equiv u$. In what follows, we adopt the $+$ sign for $r_\mathrm{bh}$ and the $-$ sign for $r_\mathrm{c}$. This coordinate system \eqref{ansatz:inEF} effectively describes ingoing waves at both $r_\mathrm{bh}$ and $r_\mathrm{c}$. To illustrate this, consider a constant $v = v_{c}$ trajectory approaching $r_\mathrm{bh}$, which corresponds to $\dd r_{*}/\dd t < 0$. However, since a single static patch of the SdS spacetime is the region $r_\mathrm{bh} < r < r_\mathrm{c}$~\cite{Galante:2023uyf}, the ingoing wave condition at $r_\mathrm{c}$ may satisfy $\dd r_{*}/\dd t > 0$. This motivates the identification $v_{-} = u$ for $r_\mathrm{c}$. 

Below, in Sections~\ref{SEC32}--\ref{SEC34}, we look for the pole-skipping points of probe scalar and Maxwell fields and of the linearized metric perturbations around the background \eqref{ORIMET}. For each of the independent fluctuations $Z(x^{\mu})$, we express the solution as 
\begin{align}\label{ANGEN}
    Z(x^{\mu}) = \bar{Z}(x^{a}) \mathbb{S}(x^{i}) ,
\end{align}
where $\bar{Z}(x^{a})$ can be further decomposed as $\bar{Z}(x^{a}) = e^{-i\omega v_{\pm}}  \bar{Z}(r)$, and $\mathbb{S}(x^{i})$ represents the spherical harmonics on the maximally symmetric space $\mathcal{K}^{d-1}$. In addition, we denote covariant derivatives on $\mathcal{N}^{2}$ by $D_a$, and those on $\mathcal{K}^{d-1}$ by $\hat{D}_i$, which will be used consistently throughout the analysis.

\paragraph{The original near-horizon method.}
To use the originally proposed near-horizon method \cite{Blake:2018leo,Blake:2019otz,Jeong:2021zhz} for finding the pole-skipping points, we consider a typical bulk differential equation 
\begin{align}\label{TOYEX}
\bar{Z}''(r) + P(r) \bar{Z}'(r) + Q(r) \bar{Z}(r) = 0 .
\end{align}
Here, the horizon $r = r_0$ (denoted as $r_\mathrm{bh}$ or $r_\mathrm{c}$ in our dS case) is a regular singularity of the differential equation. The functions $P(r)$ and $Q(r)$ admit expansions:
\begin{align}\label{PQEXP}
\begin{split}
P = \frac{P_{-1}}{r-r_0} + P_0  + \cdots , \quad
Q = \frac{Q_{-1}}{r-r_0} + Q_0 + \cdots ,
\end{split}
\end{align}
which hold for all cases considered. Typically, 
\begin{align}\label{}
\begin{split}
P_{-1} = 1 - i \frac{\omega}{2 \pi T} , \quad
Q_{-1} = Q_{-1}(\omega, k) ,
\end{split}
\end{align}
where $k$ is determined by solving the eigenvalue equations for the spherical harmonics. Note that for flat black brane horizons, $k$ is the wavevector (see e.g.~Ref.~\cite{Grozdanov:2023txs}).

We consider a Frobenius series solution as
\begin{align}
\bar{Z}(r) = \sum_{n=0}^{\infty}  \bar{Z}_{(n)} (r-r_0)^{n+\lambda} .
\end{align}
Substituting this into \eqref{TOYEX}, the indicial equation at leading order gives two independent exponents, $\lambda = 0$ and $\lambda = 1-P_{-1}$. Here, the mode with $\lambda = 0$ corresponds to the ingoing solution, while $\lambda = 1 - P_{-1} = i \omega / 2 \pi T$ represents the outgoing solution. We select the ingoing mode $\lambda = 0$.

To systematically find the pole-skipping points, we express the field equation (expanded near the horizon $r_0$) in a matrix form \cite{Blake:2018leo,Blake:2019otz,Jeong:2021zhz}:
\begin{align}
\begin{split}
M \bar{Z} 
    =
\begin{pmatrix} 
    M_{11} & M_{12} & 0 & 0 & \cdots \\
    M_{21} & M_{22} & M_{23} & 0 & \cdots \\
    \vdots & \vdots & \vdots & \vdots & \ddots 
\end{pmatrix}
\begin{pmatrix} 
    \bar{Z}_{(0)} \\  \bar{Z}_{(1)} \\  \vdots
\end{pmatrix}  
= 0,
\end{split}
\end{align}
where one can find that 
\begin{align}
M_{n, n+1} = n(n - 1 + P_{-1}) = n\left(n - i \frac{\omega}{2 \pi T}\right) .
\end{align}
The pole-skipping points are then obtained by solving
\begin{align}
M_{n,n+1}(\omega_{n}) = 0 , \quad \det \mathcal{M}^{(n)}(\omega_{n}, k_{n}) = 0 ,
\end{align}
where $\mathcal{M}^{(n)}$ is the submatrix consisting of the first $n$ rows and $n$ columns of $M$.

For instance, the leading (lowest) pole-skipping point is determined from the first row
\begin{align}\label{LEDEX}
M_{11} \bar{Z}_{(0)} + M_{12} \bar{Z}_{(1)} = Q_{-1}\bar{Z}_{(0)} + P_{-1} \bar{Z}_{(1)} =0 .
\end{align}
Normally, this equation determines the higher-order coefficient $\bar{Z}_{(1)}$ in terms of $\bar{Z}_{(0)}$. However, when $M_{12} = M_{11} = 0$ (i.e., $P_{-1} = Q_{-1} = 0$), both coefficients remain free, implying non-uniqueness in the bulk solution. This yields
\begin{align}
\begin{split}
M_{12} &= P_{-1} = 0 \quad\quad\Rightarrow\quad \omega_{1} = - i 2 \pi T , \\
\det \mathcal{M}^{(1)} &= M_{11} = Q_{-1} = 0 \quad\Rightarrow\quad k_{1} = k_{1}^{\text{model}}  ,
\end{split}
\end{align}
where $k_{1}^{\text{model}}$ is model-dependent.

Let us consider an additional example involving a subleading pole-skipping point. The field equation, expanded to the next order, takes the form
\begin{align}
\begin{split}
 M_{11} \bar{Z}_{(0)} + M_{12} \bar{Z}_{(1)} = 0 , \quad
M_{21} \bar{Z}_{(0)} + M_{22} \bar{Z}_{(1)} + M_{23} \bar{Z}_{(2)} =0 .
\end{split}
\end{align}
Rewriting these equations, we obtain
\begin{align}\label{LEDEX2}
\begin{split}
\left(M_{11}M_{22} - M_{12}M_{21}\right) \bar{Z}_{(0)}  - M_{12} M_{23} \bar{Z}_{(2)} = 0 .
\end{split}
\end{align}
It follows directly that $\bar{Z}_{(0)}$ and $\bar{Z}_{(2)}$ remain free and undetermined when the condition $M_{23}=\det \mathcal{M}^{(2)}=0$ is satisfied. This leads to the subleading pole-skipping points at
\begin{align}\label{LEDEX2}
\begin{split}
M_{23} &= 2\left[2-i{\omega}/{(2 \pi T)}\right] =0  \quad\quad\Rightarrow\quad \omega_{2} = - i 4 \pi T  \\
\det \mathcal{M}^{(2)} &= Q_{-1}(Q_{-1}+P_0)-P_{-1}Q_0 = 0 \quad\Rightarrow\quad k_{2} = k_{2}^{\text{model}} .
\end{split}
\end{align}

By extending this approach to higher orders, {\it all} pole-skipping points at higher negative imaginary multiples of the Matsubara frequency $\omega_n = - 2\pi T i n$ with $n > 0$ can in principle be  systematically derived. A potential limitation of this method is that for certain choices of the gauge-invariant variable, the method does not readily reveal the `chaotic' pole-skipping point at $\omega = + 2\pi T i$. One can either look for those points independently or use an alternative near-horizon method developed in \cite{Natsuume:2023lzy}.

\paragraph{Alternative near-horizon method.}
This alternative approach \cite{Natsuume:2023lzy}, which is also convenient when dealing with a system of fluctuation equations, reformulates \eqref{TOYEX} as a first-order matrix system:
\begin{align}\label{MEQ}
\begin{split}
\vec{X}' - M \vec{X} = 0,
\end{split}
\end{align}
where
\begin{align}\label{MATEX}
\begin{split}
\vec{X} \equiv 
\begin{pmatrix} 
\bar{Z}(r) \\ 
\bar{Z}'(r)
\end{pmatrix} , \quad
M \equiv 
\begin{pmatrix} 
  0 &1 \\
  -Q(r) & -P(r)
\end{pmatrix} .
\end{split}
\end{align}
Expanding $M$ near the horizon as
\begin{align}\label{MSERIES}
M = \frac{M_{-1}}{r-r_0} + M_0 + \cdots ,
\end{align}
the solution then again readily takes the Frobenius form
\begin{align}\label{FBE}
\vec{X}(r) = \sum_{n=0}^{\infty} \vec{X}_{(n)} (r-r_0)^{n+\lambda} .
\end{align}

Substituting this ansatz into \eqref{MEQ} and expanding it near the horizon, the leading-order condition gives an eigenvalue equation for $M_{-1}$ in \eqref{MSERIES} as
\begin{align}
\left(\lambda - M_{-1}\right) \vec{X}_{(0)} = 0 .
\end{align}
The eigenvalues and eigenvectors are
\begin{align}\label{LEXA}
\begin{split}
  & \lambda = 0 , \quad
\vec{X}_{(0)}= 
\begin{pmatrix} 
  1 \\
  x_1
\end{pmatrix} , \quad x_1= -\frac{Q_{-1}}{P_{-1}} ,
\\
  & \lambda = i  \frac{\omega}{2 \pi T} - 1 , \quad
\vec{X}_{(0)} = 
\begin{pmatrix} 
  0 \\ 
  1 
\end{pmatrix} .
\end{split}
\end{align}
We again choose the ingoing mode with $\lambda=0$.

Within this matrix formalism, one can identify special points where the coefficient $\vec{X}_{(n)}$ in \eqref{FBE} becomes `ill-defined', signaling an ambiguity in the bulk solution. In the example of Eq.~\eqref{LEXA} above, $\vec{X}_{(0)}$ is ill-defined when the coefficient $x_1$ becomes ambiguous, specifically yielding an indeterminate form $0/0$, when the conditions $P_{-1}=Q_{-1}=0$ are satisfied. This result is consistent with the leading pole-skipping point obtained via the original near-horizon method in Eq.~\eqref{LEDEX}.

The subleading pole-skipping points can also be computed in the similar way. Note that once $\vec{X}_{(0)}$ is determined, higher-order coefficient $\vec{X}_{(n)}$ can be obtained recursively through the relation
\begin{align}
\left(\lambda+n-M_{-1}\right) \vec{X}_{(n)} = \sum_{k=0}^{n-1} M_{n-1-k}  \vec{X}_{(k)} ,
\end{align}
where we consider the ingoing mode by setting $\lambda=0$. 

At the next order ($n=1$), this recursion takes the form
\begin{align}
(1-M_{-1})  \vec{X}_{(1)} = M_0  \vec{X}_{(0)} ,
\end{align}
where
\begin{align}
\vec{X}_{(1)}
  = \begin{pmatrix} 
  x_1 \\
  2x_2
    \end{pmatrix}
, \quad
x_2 = \frac{ Q_{-1}(Q_{-1}+P_0)-P_{-1}Q_0 }{ 2 P_{-1}(P_{-1}+1) } .
\end{align}
Thus, the coefficient $x_2$ becomes ill-defined when the conditions
\begin{align}
P_{-1}+1=0 , \quad
Q_{-1}(Q_{-1}+P_0)-P_{-1}Q_0=0 ,
\end{align}
are simultaneously satisfied. This corresponds to the second pole-skipping point, in agreement with the original near-horizon method result in Eq.~\eqref{LEDEX2}.

More generally, the coefficient $\vec{X}_{(n)}$ takes the form
\begin{align}
\vec{X}_{(n)}
= \begin{pmatrix} 
x_n\\
  (n+1)x_{n+1} ,
\end{pmatrix}
\end{align}
with $x_0=1$. As a result, the pole-skipping point are found when $x_n(\omega_{n},k_{n})$ have the form of $0/0$. 

As mentioned above, the use of this method does not require us to find a gauge-invariant master field or employing a master equation. Concretely, when the problem has the form of two coupled first-order differential equations for two variables, $\bar{Z}(r)$ and $\bar{Y}(r)$, the system can be represented in matrix form as  
\begin{align}\label{EOM22}
\begin{split}
\vec{X} \equiv 
\begin{pmatrix} 
\bar{Z}(r) \\ 
\bar{Y}(r)
\end{pmatrix} ,
\end{split}
\end{align}
with an appropriately defined matrix $M$. The analysis then proceeds as discussed above. 

%
\subsection{Results I: probe scalar field}\label{SEC32}
Let us first consider a minimally coupled massive probe scalar field, governed by the Klein-Gordon equation
\begin{equation}\label{Eq:KGeq}
    \left(\nabla^\mu\nabla_\mu -m_\phi^2\right)\phi=0 .
\end{equation}
To solve this equation, we adopt the ansatz \eqref{ANGEN},
\begin{equation}\label{Eq:KGeq1}
    \phi(x^{\mu}) = \bar{\phi}(x^{a})  \mathbb{S}(x^{i}) , \quad  \bar{\phi}(x^{a}) = e^{-i\omega v_{\pm}}  \bar{\phi}(r) ,
\end{equation}
where $\mathbb{S}$ represents the scalar spherical harmonics that satisfy the eigenvalue equation
\begin{equation}\label{Eq:KGeq2}
    \left(\hat{D}_i\hat{D}^i + k_S^2 \right)\mathbb{S}=0 .
\end{equation}
By substituting Eqs.~\eqref{Eq:KGeq1} and \eqref{Eq:KGeq2} into \eqref{Eq:KGeq}, we obtain the radial equation
\begin{equation}\label{Eq:scalar}
    \bar{\phi}''(r)+\left(\frac{f'(r)\mp 2i\omega}{f(r)}+\frac{d-1}{r}\right)\bar{\phi}'(r)-\frac{1}{f(r)}\left(m_\phi^2\pm\frac{i(d-1)\omega}{r}+\frac{k_S^2}{r^2}\right)\bar{\phi}(r) =0 .
\end{equation}
Comparing this with Eq.~\eqref{TOYEX}, we identify
\begin{equation}\label{}
  P(r) = \frac{f'(r)\mp 2i\omega}{f(r)}+\frac{d-1}{r}  , \quad  
  Q(r) = -\frac{1}{f(r)}\left(m_\phi^2\pm\frac{i(d-1)\omega}{r}+\frac{k_S^2}{r^2}\right)  .
\end{equation}
Expanding these functions near the horizon $r_0$ as in \eqref{PQEXP}, we obtain
\begin{align}\label{SCRE}
\begin{split}
P_{-1} = 1 \mp i \frac{2 \omega}{f'(r_0)} , \quad Q_{-1}=  -\frac{1}{f'(r_0)}\left(m_\phi^2\pm\frac{i(d-1)\omega}{r_0}+\frac{k_S^2}{r_0^2}\right) .
\end{split}
\end{align}

To determine the leading pole-skipping point with \eqref{SCRE}, we apply the condition $P_{-1}=Q_{-1}=0$, making the ill-defined coefficient $x_1$ in \eqref{LEXA}, which yields
\begin{equation}
\omega =\mp i \frac{f'(r_0)}{2}, \quad k_{S}^2=-m_{\phi}^2 r_0^2 - \frac{d-1}{2}r_0 f'(r_0) .
\end{equation}
This corresponds to the leading pole-skipping point for the scalar field.

For this pole-skipping point within the SdS black hole, we use the temperature \eqref{TSfor}. At the black hole event horizon $r_0=r_\mathrm{bh}$, choosing the negative sign and using $f'(r_\mathrm{bh})=4\pi T_\mathrm{bh}$, we obtain,
\begin{equation}
\omega_{1} = - 2 \pi  T_\mathrm{bh}i, \quad k_{S,1}^2=-m_{\phi}^2 r_\mathrm{bh}^2 - \frac{d-1}{2}r_\mathrm{bh} f'(r_\mathrm{bh}) .
\end{equation}
Similarly, at the cosmological horizon $r_0=r_\mathrm{c}$, choosing the positive sign and using $f'(r_\mathrm{c})=-4\pi T_\mathrm{c}$, we find
\begin{equation}
\label{psp:scalarcosh}
\omega_{1} = - 2\pi T_\mathrm{c} i, \quad k_{S,1}^2=-m_{\phi}^2 r_\mathrm{c}^2 - \frac{d-1}{2}r_\mathrm{c} f'(r_\mathrm{c}) .
\end{equation}
Additionally, an infinite sequence of subleading pole-skipping points appears at $\omega_n = - 2 \pi T_{\mathrm{bh}| \mathrm{c}} i n $ for $n = 2,3, \ldots$. We will not look for those subleading points here.

%
\subsection{Results II: probe Maxwell field}\label{SEC33}
We now examine the leading pole-skipping points of a Maxwell field governed by the vacuum Maxwell equations:
\begin{equation}\label{Eq:Procaeq}
    \nabla_\mu F^{\mu\nu} = 0 .
\end{equation}
At the level of linearized fluctuations, the system admits two independent modes: the (longitudinal) spin-0 scalar mode and the (transverse) spin-1 vector mode.

%
\subsubsection{Scalar mode}\label{vec:spin0}
The scalar mode is characterized by the following non-vanishing fluctuation components
\begin{equation}\label{MSM}
A_{a}(x^\mu) = \bar{A}_a(x^a)  \mathbb{S}(x^i), \quad
A_{i}(x^\mu)= \bar{A}_S(x^a)   \hat{D}_i\mathbb{S}(x^i).
\end{equation}
where $\mathbb{S}$ represents the scalar harmonics satisfying
\begin{equation}
    \begin{split}
    &\left(\hat{D}_i \hat{D}^i+k_S^2 \right)\mathbb{S}=0 .
    \end{split}
\end{equation}
Next, we introduce the gauge-invariant variable $B^a$~\cite{Kodama:2000fa, Ueda:2018xvl}:
\begin{equation}
    B^a = \bar{A}^a - D^a \bar{A}_S .
\end{equation}
The equations of motion then become
\begin{equation}
\begin{split}
    &2D_b D^{[a} B^{b]} +2(d-1)\frac{D_b r}{r}D^{[a} B^{b]}+\frac{k_S^2}{r^2}B^a  = 0,\\
    &D_b B^b + (d-3)\frac{D_b r}{r}B^b  =0,
\end{split}
\end{equation}
where only two of these equations are independent. In Fourier space, where $B^a(x^{a}) \rightarrow e^{-i\omega v_{\pm}}  B^a(r)$, we then obtain the following matrix form of first-order differential equations for $B^a(r)$:
\begin{equation}\label{eq:form2}
    \vec{X}'-M\vec{X} = 0,
\end{equation}
where
\begin{equation}
    \vec{X}=\begin{pmatrix} B^v(r) \\ B^r(r)
    \end{pmatrix}, \quad M = \begin{pmatrix}
        -\frac{f'(r)\mp 2i\omega}{f(r)} && \frac{-i k_S^2\mp(d-3) r \omega}{r^2 \omega f(r)} \\ i\omega && -\frac{d-3}{r}
    \end{pmatrix}.
\end{equation}
This is the explicit example of Eq.~\eqref{EOM22}, which can be conveniently solved by using the alternative near-horizon method.

Applying the series solution ansatz \eqref{FBE} to \eqref{eq:form2}, the leading-order eigenvalue equation reads
\begin{equation}
    (\lambda-M_{-1})\vec{X}_{(0)} =0 .
\end{equation}
The corresponding eigenvalues and eigenvectors are
\begin{equation}\label{SMRE1}
    \begin{split}
    &\lambda=0, \quad \vec{X}_{(0)}=\begin{pmatrix}-\frac{i k_S^2 \pm (d-3) r_0 \omega}{r_0^2\left(f'(r_0)\mp 2 i \omega\right)\omega} \\ 1\end{pmatrix},\\
    &\lambda = i \frac{\omega}{2\pi T}-1, \quad \vec{X}_{(0)}=\begin{pmatrix} 1\\0
    \end{pmatrix},
    \end{split}
\end{equation}
which, from the `ill-definedness' condition of $\vec{X}_{(0)}$ with $\lambda  = 0$, yields two pole-skipping points at  
\begin{equation}\label{psp:gaugeS}
\omega =0, \quad k_{S}^2=0 ,
\end{equation}
and
\begin{equation}\label{psp:gaugeS2}
\omega =\mp i \frac{f'(r_0)}{2}, \quad k_{S}^2=  \frac{d-3}{2}r_0  f'(r_0) .
\end{equation}
For the SdS black holes in any number of dimensions, these simplify to
\begin{align}
\begin{split}
&\omega_0 = 0, \quad  k_{S,0}^2=0 , \\
&\omega_{1} = - 2\pi T_{\mathrm{bh}|\mathrm{c}} i, \quad k_{S,1}^2 = \frac{d-3}{2}  r_{\mathrm{bh}|\mathrm{c}}  f'(r_{\mathrm{bh}|\mathrm{c}}) .
\end{split}
\end{align}

%
\subsubsection{Vector mode}
For the Maxwell field vector mode, the fluctuation components take the form
\begin{equation}
\label{ansatz:vectorspin1}
    A_a(x^\mu) = 0, \quad  A_{i}(x^\mu) = \bar{A}_V(x^a) \mathbb{V}_i(x^i),
\end{equation}
where the vector harmonics $\mathbb{V}_{i}$ satisfy
\begin{equation}
\begin{split}
    \left(\hat{D}_j\hat{D}^j+k_V^2 \right)\mathbb{V}_{i}=0,\quad
    \hat{D}_i\mathbb{V}^i=0.
\end{split}
\end{equation}
In Fourier space, the equation of motion simplifies to
\begin{equation}\label{Eq:gauge vector}
    \bar{A}_V''(r)+\left(\frac{f'(r) \mp 2i\omega}{f(r)}+\frac{d-3}{r}\right)\bar{A}_V'(r)-\frac{1}{f(r)}\left(\pm\frac{i(d-3)\omega}{r}+\frac{k_V^2+(d-2)}{r^2}\right)\bar{A}_V(r) = 0 .
\end{equation}

We can now determine the leading pole-skipping point for the SdS black holes by identifying the  functions $P(r)$ and $Q(r)$ and solving $P_{-1}=Q_{-1}=0$ in \eqref{LEXA}. This yields
\begin{equation}\label{eq:fpspProcaV}
        \omega_{1} = - 2\pi T_{\mathrm{bh}|\mathrm{c}} i , \quad k_{V, 1}^2=-(d-2) - \frac{d-3}{2} r_{\mathrm{bh}|\mathrm{c}} f'(r_{\mathrm{bh}|\mathrm{c}}).
\end{equation}

%
\subsection{Results III: gravitational field}\label{SEC34}
Finally, we examine pole-skipping phenomenon for the gravitational field, governed by the Einstein's equation
\begin{equation}\label{Eq:EE}
    G_{\mu\nu} + \Lambda g_{\mu\nu} = 8\pi G_N T_{\mu\nu} ,
\end{equation}
where $G_{\mu\nu}$ is the Einstein tensor. Here, the linearized metric fluctuations can be decomposed into three independent channels: the spin-0 sound modes, the spin-1 shear modes, and the spin-2 tensor modes. We analyze their corresponding leading pole-skipping points separately.

%
\subsubsection{Sound mode}
The gravitational sound mode requires the following nonvanishing components of the linearized metric perturbations:
\begin{align}
\begin{split}
    \delta g_{ab}(x^{\mu}) &= \bar{g}_{ab}(x^{a})  \mathbb{S}(x^{i}) , \\  
   \delta g_{ai}(x^{\mu})  &= r  \bar{g}_{a}(x^{a})  \mathbb{S}_i(x^{i}) , \\  
    \delta g_{ij}(x^{\mu})   &= 2  r^2 \left[\bar{g}_{L}(x^{a})  \gamma_{ij}\mathbb{S}(x^{i}) + \bar{g}_{T}(x^{a})  \mathbb{S}_{ij}(x^{i}) \right] ,
\end{split}
\end{align}
where the scalar harmonics $\mathbb{S}$ satisfy
\begin{align}
\begin{split}
    \left(\hat{D}_i \hat{D}^i+k_S^2 \right)\mathbb{S} = 0, 
\end{split}
\end{align}
with additional relations
\begin{align}
\begin{split}
    \mathbb{S}_i = -\frac{1}{k_S}\hat{D}_i\mathbb{S},\quad
    \mathbb{S}_{ij} = \frac{1}{k_S^2}\hat{D}_i\hat{D}_j\mathbb{S}+\frac{1}{d-1}\gamma_{ij}\mathbb{S}.
\end{split}
\end{align}

To analyze pole-skipping, we introduce the gauge-invariant variable~\cite{Kodama:2000fa, Ueda:2018xvl}
\begin{equation}
\begin{split}
    H_a &= \frac{r}{k_S}\left(\bar{g}_a + \frac{r}{k_S} D_a  \bar{g}_T\right),\\
    F &= \bar{g}_L+\frac{1}{d-1} \bar{g}_T+\frac{1}{r}D^a r H_a,\\
    F_{ab} &= \bar{g}_{ab}+D_a H_b +D_b H_a.
\end{split}
\end{equation}
In Fourier space, the independent equations take the form of a first-order system
\begin{equation}\label{eq:form3}
    \vec{X}'-M\vec{X} =0,
\end{equation}
where $\vec{X}=\left\{F(r), F_{vv}(r)\right\}$, and $M$ is a $2 \times 2$ matrix with complex elements. For brevity, we omit the explicit form of $M$.

\begin{figure}[]
  \centering
     {\includegraphics[width=0.47 \linewidth]{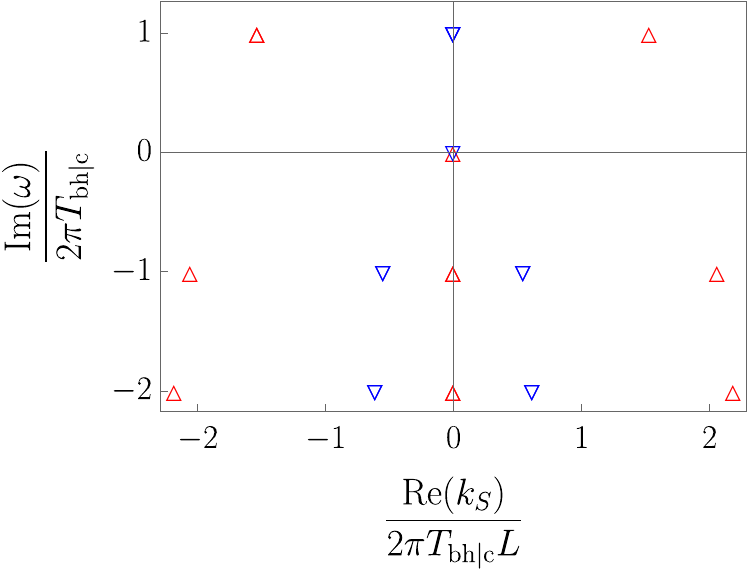} \label{}}
\quad
     {\includegraphics[width=0.47 \linewidth]{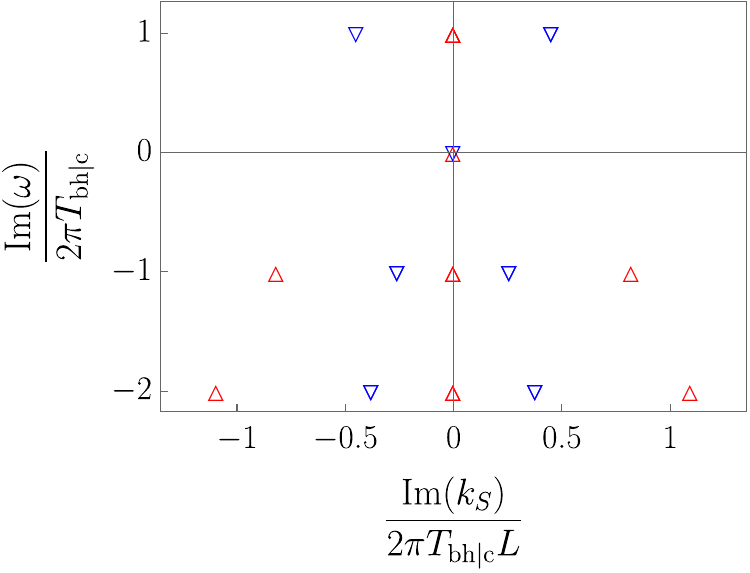} \label{}}
 \caption{Pole-skipping points of the gravitational sound mode in the four-dimensional SdS black hole geometry with $M/L=1/10$. Red data denote the pole-skipping points computed at the cosmological horizon, while blue data correspond to those computed at the black hole horizon.}\label{fig:pspmetricsound}
\end{figure}

Applying the series expansion \eqref{FBE} to \eqref{eq:form3}, the leading-order equation takes the form
\begin{equation}
    \left(\lambda-M_{-1}\right)\vec{X}_{(0)} = 0 .
\end{equation}
The corresponding eigenvalues and eigenvectors are
\begin{equation}
\begin{split}
    &\lambda = 0, \quad \vec{X}_{(0)} =\begin{pmatrix}\mp i\frac{r_0 \omega \pm i k_S^2/(d-1)}{2r_0^2 \omega(2\pi i T - \omega)}\\1
    \end{pmatrix},\\
    &\lambda = i \frac{\omega}{2\pi T},\quad \vec{X}_{(0)} = \begin{pmatrix}\mp i\frac{r_0 \omega \mp i k_S^2/(d-1)}{2r_0^2 \omega(2\pi i T + \omega)}\\1
    \end{pmatrix}.
\end{split}
\end{equation}
From the singular structure of $\vec{X}{(0)}$, we identify three pole-skipping points. For the SdS black hole, they are given by
\begin{equation}\label{MPSP}
    \begin{split}
        &\omega_{-1} =   2\pi T_{\mathrm{bh}|\mathrm{c}} i, \quad k_{S, -1}^2=\mp2(d-1)\pi  r_{\mathrm{bh}|\mathrm{c}} T_{\mathrm{bh}|\mathrm{c}} ,\\
        &\omega_{0} = 0 , \quad  k_{S, 0}^2=0,\\
        &\omega_{1} = -2 \pi T_{\mathrm{bh}|\mathrm{c}} i , \quad k_{S,1}^2=\begin{cases}(d-2)\mp2 \pi  (3-d) r_{\mathrm{bh}|\mathrm{c}} T_{\mathrm{bh}|\mathrm{c}}+h\\ (d-2)\mp2 \pi  (3-d) r_{\mathrm{bh}|\mathrm{c}} T_{\mathrm{bh}|\mathrm{c}}-h\end{cases},\\
        &\text{where}\qquad h \equiv \sqrt{(d-1) (d-2)-(d-2) (1\pm 4 \pi r_{\mathrm{bh}|\mathrm{c}} T_{\mathrm{bh}|\mathrm{c}})^2}\,.
    \end{split}
\end{equation}
Here, the negative sign in $k_S^2$ corresponds to the black hole horizon, while the positive sign corresponds to the cosmological horizon. Notably, we find both the `chaotic' pole-skipping point ($\omega_{-1} = 2\pi T_{\mathrm{bh}|\mathrm{c}} i$) and the `hydrodynamic' pole-skipping point ($\omega_{0} = 0$).\footnote{Our leading pole-skipping point in Eq.~\eqref{MPSP} is consistent with findings previously reported in the literature, specifically aligning with the results of \cite{Grozdanov:2023txs} for the case $d=3$, upon making the appropriate identification of variables, such as $k_S^2 = \mu$.} In Figure~\ref{fig:pspmetricsound}, we visualize the structure of the pole-skipping points including \eqref{MPSP}. In the next section, we will investigate the chaotic pole-skipping point in \eqref{MPSP} and its potential connection to the butterfly velocity of a hypothetical dual theory.

%
\subsubsection{Shear mode}
For gravitational shear modes, the non-vanishing metric fluctuations take the form
\begin{equation}
     \delta g_{ai}(x^{\mu}) = r  \bar{g}_a(x^{a})  \mathbb{V}_i(x^{i}) , \quad 
     \delta g_{ij}(x^{\mu}) = 2  r^2  \bar{g}_{T}(x^{a})  \mathbb{V}_{ij}(x^{i}) .
\end{equation}
Here, $\mathbb{V}_i$ represents the vector harmonics, satisfying
\begin{equation}
\begin{split}
    \left(\hat{D}_j\hat{D}^j+k_V^2 \right)\mathbb{V}_{i}=0.
\end{split}
\end{equation}
Additionally, we introduce a symmetric and traceless tensor $\mathbb{V}_{ij}$, defined as
\begin{equation}
    \mathbb{V}_{ij}=-\frac{1}{k_V}\left(\hat{D}_i\mathbb{V}_j+\hat{D}_j\mathbb{V}_i\right).
\end{equation}
A gauge-invariant combination can be constructed as
\begin{equation}
    F_a = \bar{g}_a+\frac{r}{k_V}  D_a \bar{g}_T .
\end{equation}
In Fourier space, this leads to the first-order equation
\begin{equation}
    \vec{X}'-M\vec{X} =0,
\end{equation}
where
\begin{equation}
    \vec{X}=\begin{pmatrix} F_v \\ F_r
    \end{pmatrix}, \quad M = \begin{pmatrix}
         M_{vv}&& M_{vr} \\ M_{rv} && M_{rr}
    \end{pmatrix}.
\end{equation}
The matrix components are given by
\begin{equation}
\begin{split}
&M_{vv}=\frac{1}{r}\pm\frac{i \left(k_V^2-(d-2)\right)}{r^2 \omega },\\
&M_{vr}=\frac{i f(r) \left(k_V^2-(d-2)\right)}{r^2 \omega }-i \omega,\\
&M_{rv}=-\frac{1}{f(r)}\left(\frac{i \left(k_V^2-(d-2)\right)}{r^2 \omega }\mp\frac{1-d}{r}\right),\\
&M_{rr}=\frac{2-d}{r}\mp\frac{i \left(k_V^2-(d-2)\right)}{r^2 \omega }-\frac{f'(r)\mp2 i \omega }{f(r)}.
\end{split}
\end{equation}

Applying a series expansion to this system, the leading-order equation takes the form
\begin{equation}
    \left(\lambda-M_{-1}\right)\vec{X}_{(0)}=0 ,
\end{equation}
with the eigenvalues and eigenvectors:
\begin{equation}
    \begin{split}
    &\lambda=0, \quad\vec{X}_{(0)}=\begin{pmatrix}i\frac{r_0^2 \omega  (f'(r_0) \mp 2 i \omega  )}{k_V^2-(d-2) \mp i (d-1) r_0 \omega} \\ 1\end{pmatrix},\\
    &\lambda = \pm i \frac{\omega}{2\pi T}-1, \quad \vec{X}_{(0)}=\begin{pmatrix} 0\\1
    \end{pmatrix}.
    \end{split}
\end{equation}
The pole-skipping points are then given by
\begin{equation}
    \begin{split}
        &\omega_0=0, \quad\quad\quad k_{V,0}^2 = (d-2),\\
        &\omega_{1}=-i  2\pi T_{\mathrm{bh}|\mathrm{c}}, \quad k_{V,1}^2=(d-2) \pm 2 (d-1)\pi  r_{\mathrm{bh}|\mathrm{c}} T_{\mathrm{bh}|\mathrm{c}} .
    \end{split}
\end{equation}
Here, the positive sign in $k_{V,1}^2$ corresponds to the black hole horizon ($r_\mathrm{bh}$), while the negative sign corresponds to the cosmological horizon ($r_\mathrm{c}$).

\subsubsection{Tensor mode}
Finally, the gravitational tensor modes exist for $d>3$ and are expressed with the tensor harmonics $\mathbb{T}_{ij}$, which satisfy
\begin{equation}
    \begin{split}
    &\left(\hat{D}_k \hat{D}^{k}+k_T^2 \right)\mathbb{T}_{ij}=0,\\
    &\mathbb{T}_i^i=0, \quad \hat{D}_j\mathbb{T}_i^j=0.
    \end{split}
\end{equation}
Using these harmonics, the non-vanishing metric fluctuation is
\begin{equation}
    \delta g_{ij}(x^{\mu}) = 2  r^2  \bar{g}_{T}(x^{a}) \mathbb{T}_{ij}(x^{i}).
\end{equation}
In Fourier space, the equation of motion for $\bar{g}_T(r)$ is
\begin{equation}
\label{Eq:metric tensor}
    \bar{g}_T''(r)+\left(\frac{f'(r)\mp 2i\omega}{f(r)}+\frac{d-1}{r}\right)\bar{g}_T'(r)-\frac{1}{f(r)}\left(\frac{k_T^2+2}{r^2}\pm\frac{i(d-1)\omega}{r}\right)\bar{g}_T(r) = 0.
\end{equation}
Applying the single-value formalism \eqref{TOYEX}, the pole-skipping point for SdS black holes is determined as
\begin{equation}\label{eq:fpspmetricT}
        \omega_{1} = -i  2\pi T_{\mathrm{bh}|\mathrm{c}} , \quad k_{T, 1}^2 = -2\mp 2(d-1)\pi  r_{\mathrm{bh}|\mathrm{c}} T_{\mathrm{bh}|\mathrm{c}}.
\end{equation}
The negative sign in $k_{T,1}^2$ corresponds to $r_\mathrm{bh}$, while the positive sign corresponds to $r_\mathrm{c}$.

%
\section{The hypothetical butterfly velocities in de Sitter spacetimes}\label{sec4}

%
\subsection{Quantum chaos and scrambling in holography: a review}

Quantum chaos has proven invaluable for understanding the dynamics of gravitational horizons, while horizon physics itself has shed new light on quantum chaotic phenomena. Notably, black holes in Anti-de Sitter (AdS) space have drawn special attention due to their holographic interpretation in terms of dual quantum field theories. Although a complete holographic framework for de Sitter (dS) space remains elusive, insights from AdS/CFT suggest potential pathways for discovering dS duals and their defining properties. Here, we briefly review our current understanding of quantum chaos in both AdS and dS spacetimes.

\paragraph{Out-of-time-ordered correlators.}
A common way to diagnose chaos in quantum many-body systems is to examine how an initial perturbation $V$ influences another operator $W$ measured at later time. This effect is captured by the squared commutator~\cite{Larkin1969QuasiclassicalMI}:  
\begin{equation}\label{eq:commutator}
C(t) = -\langle [W(t),V(0)]^2 \rangle,
\end{equation}  
where \( \langle\cdots\rangle = Z^{-1} \Tr [e^{\beta H} \cdots] \) denotes the thermal expectation value at temperature \( T = \beta^{-1} \). The characteristic time at which \( C(t) \) becomes significant is known as the scrambling time, \( t_* \).  
The time-ordered contribution to $C(t)$, 
\begin{equation}
\langle V(0) V(0) W(t) W(t) \rangle \sim \langle V V \rangle \langle W W \rangle + \mathcal{O}(e^{-t/t_{\text{th}}}),
\end{equation}  
decays on a timescale known as the thermalization time \( t_{\text{th}} \). In holographic systems dual to AdS black holes, this timescale corresponds to the inverse of the lowest quasinormal mode and is typically of the order \( t_{\text{th}} \sim \beta \ll t_*\). Instead, chaotic behavior can be diagnosed through the out-of-time-ordered correlator (OTOC),
\begin{equation}
f(t)=\frac{\langle V(0) W(t) V(0) W(t) \rangle}{\langle V V \rangle \langle W W \rangle},
\end{equation}  
which captures the exponential sensitivity of the system to initial conditions. In particular, this observable remains `large' for a long period of time if the system is chaotic. More concretely, in holographic large-$N$ theories dual to classical AdS black holes, one finds that for $t_{\text{th}}\ll t \ll t_*$, 
\begin{equation} \label{eq:OTOChomo}
f(t)=1-\frac{f_0}{N}e^{\lambda_L t}+\mathcal{O}(N^{-2}),
\end{equation}
where \( N \) is the number of degrees of freedom of the theory, \( f_0 \) is an $\mathcal{O}(1)$ constant that depends on the particular choice of $V$ and $W$, and \( \lambda_L=2\pi /\beta \) is the so-called quantum Lyapunov exponent. The time at which \( f(t) \) becomes negligible is precisely the time at which \( C(t) \) becomes significant, i.e., the scrambling time,  
\(t_*\sim \beta \log N\). In systems with a small number of `local' degrees of freedom $N$ (for example, in quantum spin chains), instead, OTOCs saturate at early times and, instead, chaos can be diagnosed by polynomially growing OTOCs of spatially smeared operators \cite{Kukuljan:2017xag}.

In theories with late-time exponential growth \eqref{eq:OTOChomo}, the quantum Lyapunov exponent \( \lambda_L \) has been proven to satisfy a universal bound on chaos, applicable to any quantum system with a well-defined semiclassical limit. This bound is given by~\cite{Maldacena:2015waa}  
\begin{equation}\label{eq:boundLambda}
\lambda_L \leq \frac{2\pi}{\beta},
\end{equation}  
and is thus saturated by large-\( N \) theories dual to classical black holes in AdS space. This result provides strong evidence for the conjecture that black holes are among the fastest scramblers in nature~\cite{Sekino:2008he}. Furthermore, it has served as a robust diagnostic criterion for identifying theories with possible holographic duals, as the bound is precisely saturated by certain conformal field theories (CFTs) with large central charge $c$~\cite{Roberts:2014ifa, Perlmutter:2016pkf} and by variants of the Sachdev-Ye-Kitaev (SYK) model, in the context of AdS$_2$ holography~\cite{Maldacena:2016hyu,Sarosi:2017ykf}.\footnote{Certain correlators that are subleading in the $1/N$ expansion also saturate the bound, a phenomenon that can be attributed to closed-open string duality \cite{deBoer:2017xdk, Murata:2017rbp, Banerjee:2018twd, Banerjee:2018kwy}.}

A refinement of the above observable, which is sensitive to the spatial spread of quantum chaos, involves considering perturbations \( W(t,\vec{x}) \). In this case, the relevant OTOC takes the form  
\begin{align}\label{OTOCBEHA}
\begin{split}
f(t,\vec{x})=\frac{\langle V(0) W(t,\vec{x}) V(0) W(t,\vec{x}) \rangle}{\langle V V \rangle \langle W W \rangle}.
\end{split}    
\end{align}
For theories exhibiting quantum chaos, Eq.~(\ref{eq:OTOChomo}) generalizes to  
\begin{equation} \label{eq:OTOChomoX}
f(t,\vec{x})=1-\frac{f_0}{N}e^{\lambda_L \left(t-\frac{|\vec{x}|}{v_B}\right)}+\mathcal{O}(N^{-2}), \quad \text{for }|\vec{x}|\gg\beta.
\end{equation}
The parameter $v_B$, known as the butterfly velocity, quantifies the rate at which the operator $W$ spreads spatially. This velocity defines an emergent causal structure, delineated by the butterfly light cone, given by \( t - t_* = |\vec{x}|/v_B \). Within this causal region, i.e., for \( t - t_* > |\vec{x}|/v_B \), the squared commutator behaves as \( C(t,\vec{x}) \sim \mathcal{O}(1) \), whereas outside the cone, for \( t - t_* < |\vec{x}|/v_B \), one finds \( C(t,\vec{x}) \approx 0 \). Notably, it was argued in~\cite{Roberts:2016wdl} that \( v_B \) serves as an effective low-energy Lieb-Robinson velocity, imposing a fundamental bound on the rate of quantum information propagation. 

In~\cite{Mezei:2016zxg}, it was proven that, under certain simplifying assumptions, the butterfly velocity for holographic systems satisfies the bound  
\begin{equation} \label{boundvB}
v_B\leq v_B^{\text{Sch}}=\sqrt{\frac{d}{2(d-1)}},
\end{equation}
where \( v_B^\text{Sch} \) is the butterfly velocity for the theory dual to a \((d+1)\)-dimensional AdS-Schwarzschild black hole. The key assumptions underlying this result include spatial homogeneity and isotropy, two-derivative (Einstein) gravity with asymptotically AdS boundary conditions, and bulk matter obeying the null energy condition (NEC).  Naturally, it was tempting to conjecture that Eq.~(\ref{boundvB}) might serve as a universal bound for any quantum system, analogous to the bound on the Lyapunov exponent in Eq.~(\ref{eq:boundLambda}). However, (\ref{boundvB}) was shown to fail in theories with higher-derivative (stringy) corrections~\cite{Roberts:2014isa,Alishahiha:2016cjk,Grozdanov:2018kkt,Dong:2022ucb}, as well as in anisotropic systems within Einstein gravity~\cite{Giataganas:2017koz,Jahnke:2017iwi,Gursoy:2020kjd}, reminiscent of the well-known violations of the shear viscosity-to-entropy density ratio ($\eta/s$) in such systems~\cite{Kats:2007mq,Brigante:2007nu,Brigante:2008gz,Camanho:2010ru,Erdmenger:2010xm,Rebhan:2011vd,Jahnke:2014vwa}. The correlation between the decrease of $v_B$ and increase of $\eta/s$ in the presence of certain higher-derivative corrections was discussed in Ref.~\cite{Grozdanov:2018kkt}. 

Nevertheless, even in the cases mentioned above, the butterfly velocity remains bounded from above and does not reach the speed of light \( c = 1 \), provided that the theory preserves causality. However, in scenarios where causality is violated, $v_B$ can exceed the speed of light. For instance, in Einstein-Gauss-Bonnet gravity in \( d=4 \) dimensions (see e.g.~Ref.~\cite{Grozdanov:2016fkt}), one finds \( v_B > 1 \) for \(\lambda_{\text{GB}} < -0.75 \), while causality constraints require \(\lambda_{\text{GB}} > -0.19\)~\cite{Camanho:2009vw,Buchel:2009sk}. Naively, one might expect the speed of light to define the maximal velocity of causal influence in a relativistic system. However, as clarified in~\cite{Qi:2017ttv}, when access is restricted to a subset of the Hilbert space, the effective causal propagation velocity can be generically smaller than the speed of light. This explains why \( v_B < 1 \) in typical cases, as $v_B$ governs causal propagation within the subset of the Hilbert space associated with the thermal ensemble. Indeed, the authors of~\cite{Qi:2017ttv} demonstrated that, for any asymptotically AdS geometry described by two-derivative gravity, the butterfly velocity is strictly bounded by the speed of light:
\begin{equation}\label{boundvB2}
v_B \leq 1,
\end{equation}
as expected for a theory with a Lorentz-invariant UV fixed point.

Anticipating our analysis for the dS case, it is worth highlighting certain results on the butterfly velocity for holographic setups beyond the standard AdS/CFT framework. Indeed, in going beyond asymptotically AdS backgrounds, one finds examples in top-down non-AdS/non-CFT dualities where the butterfly velocity can exceed the speed of light due to nonlocal interactions in the dual field theory. Concretely, this phenomenon has been observed in the gravity duals of Non-Commutative Super Yang-Mills (NCSYM) and Dipole Deformed Super Yang-Mills (DDSYM)~\cite{Fischler:2018kwt,Eccles:2021zum}, provided the degree of nonlocality is large compared to the characteristic thermal length scale \(\beta\). Furthermore, the same effect has been observed in (non-holographic) lattice models that incorporate spatial nonlocality~\cite{Eccles:2021zum}, indicating that this feature is not exclusive to holographic systems. In all of these theories, standard constraints from local quantum field theories are explicitly relaxed, thus allowing superluminal signal propagation. Consequently, it is not surprising that these models---which violate causality by design through the inclusion of sufficiently long-range interactions---admit a butterfly velocity exceeding the speed of light in certain regimes.

\paragraph{Bulk scattering and shock wave analysis.} In holographic theories, the study of OTOCs can be reformulated in terms of a high-energy scattering process near the black hole horizon~\cite{Shenker:2013pqa,Shenker:2013yza,Shenker:2014cwa,Roberts:2014isa}. Specifically, one views the operators in the OTOC as creating excitations that scatter off each other in the bulk, with the near-horizon region playing a crucial role. In the Regge limit (large center-of-mass energy and small momentum transfer), the scattering amplitude is dominated by the eikonal approximation, where ladder diagrams in the bulk resum into an exponential (the eikonal phase), which manifests in the bulk as a shock wave geometry~\cite{Dray:1984ha,tHooft:1987vrq,Cornalba:2006xk}.

Concretely, consider two boundary insertions separated by a large time interval. In the bulk, these insertions become wavepackets traveling in opposite directions. Their collision, at high center-of-mass energy, occurs close to the horizon, where the geometry can be approximated by an AdS-Rindler background. The subsequent amplitude for this near-horizon scattering captures the essence of chaos as encoded by the OTOC. The key simplification in this regime is that the large energy of one wavepacket backreacts on the geometry, producing a shock wave profile~\cite{Dray:1984ha,tHooft:1987vrq}. In the eikonal limit, such a shock wave modifies the horizon by introducing a shift in the light-cone coordinate. Physically, this deformation encodes how the early-time perturbation dramatically influences late-time observables. Gravitational shock wave solutions were initially studied in flat space but extend naturally to AdS or even dS space~\cite{Hotta:1992qy,Hotta:1992wb,Sfetsos:1994xa}, and are thus expected to be robust in the study of quantum chaotic properties of general spacetimes.

In practice, one solves the linearized Einstein equations around a black hole background with a null source localized on the horizon. The resulting metric perturbation represents a shock wave traveling at the speed of light along the horizon. Holographically, the backreacted profile of the shock wave in the bulk translates into an exponential suppression of OTOCs, from which one may extract the corresponding Lyapunov exponent \(\lambda_L\) and butterfly velocity \(v_B\), confirming the link between the bulk scattering amplitude and the chaotic features of the black hole horizon~\cite{Shenker:2014cwa,Roberts:2014ifa}.

This method has been corroborated by direct calculations in simple boundary QFTs. Examples include the BTZ black holes dual to two-dimensional large-$c$ CFTs~\cite{Shenker:2014cwa,Roberts:2014ifa,Poojary:2018esz,Jahnke:2019gxr,Cotler:2018zff,Haehl:2018izb}, AdS\(_2\) gravity and their SYK-like quantum mechanical duals~\cite{Kitaev-2014,Maldacena:2016hyu,Jensen:2016pah,Maldacena:2016upp,Engelsoy:2016xyb}, and hyperbolic AdS black holes dual to CFTs in hyperbolic space~\cite{Perlmutter:2016pkf,Ahn:2019rnq}. Taken together, these analyses show that this approach not only reproduces the chaotic features diagnosed by OTOCs but also provides a clear geometric intuition: early-time perturbations dramatically perturb the horizon via shock waves, whose backreaction encodes the effective scrambling dynamics in strongly coupled holographic theories.

\paragraph{Pole-skipping and horizon constraints.}
Recent developments have demonstrated that quantum chaos can also be probed via the analytic structure of energy density two-point correlation functions~\cite{Grozdanov:2017ajz,Blake:2017ris}. This insight complements earlier studies involving OTOCs and gravitational shock wave analyses while, unlike in the case of OTOCs, working directly with unambiguously physically observable quantities. A distinctive signature of quantum chaos in this context is the phenomenon of \textit{pole-skipping}, first discovered in a maximally chaotic holographic model~\cite{Grozdanov:2017ajz} and then established as a universal feature of maximally chaotic systems through effective field theory arguments~\cite{Blake:2017ris}.\footnote{See also the recent development~\cite{Chua:2025vig} where pole-skipping has been precisely associated with the gravitational replica manifold for the late-time entanglement wedge and the shock wave geometry.}

Pole-skipping occurs at specific points in momentum space where the energy density two-point function becomes non-single-valued due to a simultaneous vanishing of both the pole and its residue---formally, a `$0/0$' in the correlator. The special chaotic pole-skipping point in the upper-half of the complex $\omega$ plane is characterized by
\begin{align}\label{ADSPSC}
\omega_{*} = i \lambda_L , \quad k_{*} = i \frac{\lambda_L}{v_B} \quad\Longrightarrow\quad v_B^2=\frac{\omega_{*}^2}{k_{*}^2} ,
\end{align}
where $\lambda_L$ and $v_B$ are the quantum Lyapunov exponent and butterfly velocity, respectively. At this momentum-space location, two independent solutions emerge at the horizon, breaking the uniqueness required to define the boundary correlator. In terms of the holographic description, pole-skipping arises purely from the near-horizon physics rather than requiring a full solution of the bulk wave equations. 

The connection between pole-skipping in two-point functions of energy-energy correlators and OTOCs closely follows from the underlying link between hydrodynamics and chaos. A simple way to motivate it is by expressing an OTOC in terms of plane waves proportional to $\exp (-i \omega t + i k x )$, evaluated precisely at the pole-skipping frequency and wave number $(\omega,k)=(\omega_{*},k_{*})$. In such cases, the butterfly velocity $v_B$ emerges naturally, reflecting how rapidly perturbations spread in chaotic quantum systems.

Pole-skipping has been extensively validated across diverse holographic settings, including massive gravity models breaking translational symmetry~\cite{Blake:2018leo,Jeong:2021zhz}, SYK chains~\cite{Gu:2016oyy,Choi:2020tdj}, two-dimensional large-$c$ CFTs~\cite{Haehl:2018izb}, $d$-dimensional CFTs in hyperbolic spaces~\cite{Ahn:2020bks,Ahn:2020baf}, and more recently, charged JT gravity coupled to complex SYK models~\cite{Yuan:2023tft}. These examples underscore pole-skipping's robustness as a diagnostic tool for quantum chaos and highlight its universality within holographic frameworks.

\paragraph{Quantum chaos and scrambling in de Sitter space.} Extensive studies of quantum chaos and scrambling in AdS black holes—through OTOCs and pole-skipping, along with the details of their gravitational bulk descriptions (non-linear shock waves and linearized fluctuations, respectively)—naturally motivate similar investigations in dS black holes within the framework of dS holography. 

Historically, the presence of a cosmological horizon in dS space, along with the associated Rindler-like behavior near it, led to the conjecture that dS space is also a `fast scrambler'~\cite{Susskind:2011ap,Geng:2020kxh,Blommaert:2020tht}. Specifically, it has been proposed that the scrambling time scales as
\begin{equation}\label{}
        t_{*}\beta\approx\log S_{\text{GH}} ,
\end{equation}
where $S_{\text{GH}}$ is the Gibbons-Hawking entropy, similar to the behavior in black holes.

In the context of dS black holes, studies of OTOCs  have primarily focused on the Lyapunov exponent, often without considering spatial profiles. Some key developments in this direction include:
(I) Shockwave computations in three-dimensional dS~\cite{Aalsma:2020aib}, which yield $\lambda_L=2\pi/\beta$, despite an ‘inverse’ Gao-Wald effect.\footnote{Unlike in Minkowski and AdS space, where geodesics experience time delays when crossing positive-energy shockwaves, geodesics in dS undergo time advances \cite{Gao:2000ga}. This suggests that perturbations satisfying the null energy condition can facilitate signal propagation between otherwise causally disconnected regions, akin to traversable wormholes in AdS~\cite{Gao:2016bin,Maldacena:2017axo}.} Moreover, it was found that OTOCs in dS develop oscillations \textit{after} $t_*$, even though the single-sided OTOC still saturates the chaos bound. 
(II) The hyperfast scrambling conjecture~\cite{Susskind:2021esx}, which argues that the rapid convergence of spacelike geodesics to future and past infinities in dS could imply that the holographic observables diverge at a finite time. This notion has been further explored in the double-scaled Sachdev-Ye-Kitaev (DSSYK) model~\cite{Berkooz:2018jqr} and has been the subject of several follow-up investigations~\cite{Susskind:2022dfz,Lin:2022nss,Lin:2022rbf,Susskind:2022bia,Rahman:2022jsf,Narovlansky:2023lfz,Verlinde:2024znh,Bhattacharjee:2022ave,Goel:2023svz,Rabinovici:2023yex,Blommaert:2023opb,Ambrosini:2024sre,Xu:2024hoc,Xu:2024gfm,Heller:2024ldz}.\footnote{For discussions on chaos and holographic complexity in dS space, see also~\cite{Chapman:2021eyy,Reynolds:2017lwq,Jorstad:2022mls,Galante:2022nhj,Auzzi:2023qbm,Anegawa:2023wrk,Anegawa:2023dad,Baiguera:2023tpt,Aguilar-Gutierrez:2023zqm,Aguilar-Gutierrez:2023tic,Aguilar-Gutierrez:2024rka,Aguilar-Gutierrez:2024nau}.}
(III) Holographic OTOC computations in two-dimensional gravitational setups~\cite{Anninos:2018svg}, where a dilaton-gravity model flows from an asymptotically AdS$_2$ region to dS$_2$ in the deep interior. In this scenario, OTOCs oscillate in time rather than grow exponentially, with the oscillation frequency matching $\lambda_L$.\footnote{The OTOC computed in this work was for massless perturbations. Massless fields might behave qualitatively differently, since there is no vacuum state for a massless scalar field in de Sitter space that is invariant under the full isometry group \cite{Allen:1987tz}.} Notably, a related two-dimensional analysis~\cite{Kolchmeyer:2024fly} found that $\lambda_L$ differs by a factor of two from the three-dimensional result~\cite{Aalsma:2020aib}, a discrepancy attributed to differences in geometric considerations.\footnote{In \cite{Kolchmeyer:2024fly}, OTOCs are computed using a dynamical observer, whereas in \cite{Aalsma:2020aib}, they are calculated with a dynamical metric.}

These findings indicate that while chaos and scrambling remain relevant in dS space, their specific characteristics exhibit similarities and also differences when compared to black holes in AdS/CFT.

\paragraph{Scope of this work.} In this section, we extend these investigations in several directions: (I) We generalize the analysis of OTOCs to dS space and SdS black holes in $d\geq3$ dimensions.\footnote{In $d=1$, Einstein gravity is topological. One may find dS solutions and SdS black holes in dilaton gravities, but they lie outside the scope of our work. In $d=2$, classical SdS black holes do not exist, instead, the solutions represent conical defects. However, semiclassical backreaction allows for true 3d SdS black holes \cite{Emparan:2022ijy}. Likewise, these solutions are outside the scope of our work.
}
(II) We examine whether the Lyapunov exponent $\lambda_L$ and butterfly velocity $v_B$ can be consistently determined from both the pole-skipping and shock wave analyses. In other words, we ask if these two approaches are compatible and whether they provide a meaningful characterization of chaos in asymptotically dS space and dS black holes. 
(III) Lastly, we discuss a possible microscopic interpretation of our results in the context of DSSYK-like chain models and their implications for chaotic dynamics.\footnote{For recent discussions on the connection between pole-skipping in two-dimensional dS spacetime and DSSYK—particularly for scalar and fermionic fields—see~\cite{Yuan:2024utc}.}

%
\subsection{Pole-skipping of the gravitational sound mode}\label{sec42}
\begin{figure}[]
  \centering
     {\includegraphics[width=7.2cm]{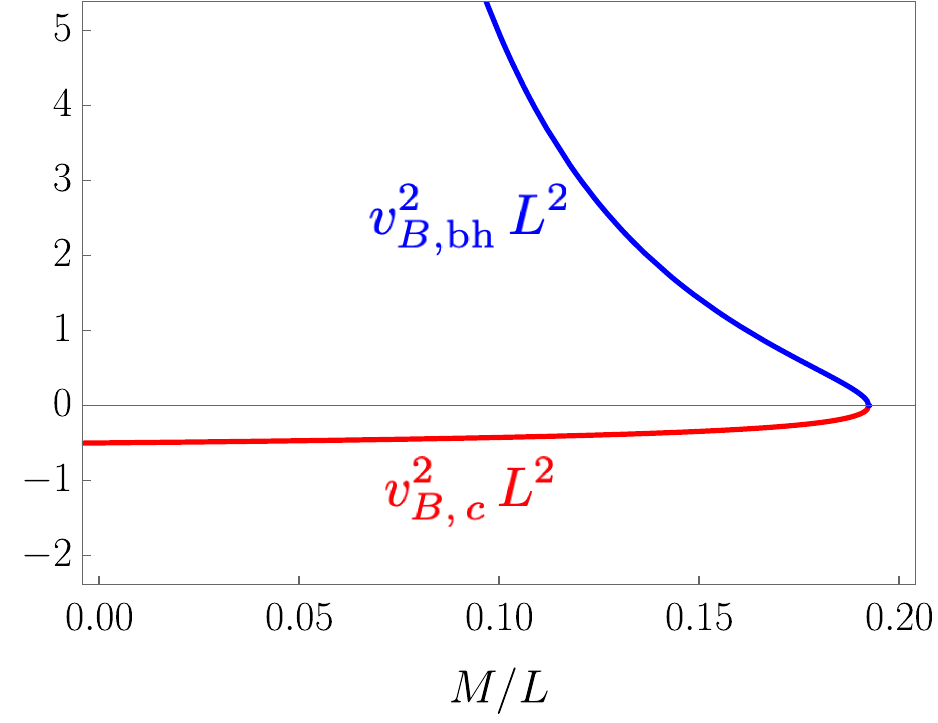} \label{}}
 \caption{Butterfly velocities of SdS as given in Eq.~\eqref{vbfor} for $d=3$, which reduce to \eqref{vbfor2} in the small $M/L$ limit.}\label{fig:masshorizonvB}
\end{figure}

If we adopt the pole-skipping  phenomenon of the gravitational sound mode (in the upper-half complex $\omega$ plane) as a `working definition' of the butterfly effect, then our results for SdS black holes \eqref{MPSP} imply that
\begin{equation}\label{VBRESULT2}
    \lambda_L \equiv-i\omega_{-1}= 2\pi T_{\mathrm{bh}|\mathrm{c}},
\end{equation}
and
\begin{equation}\label{VBRESULT}
    v_{B,\mathrm{bh}|\mathrm{c}}^2 \equiv\frac{\omega_{-1}^2}{k_{S,-1}^2} .
\end{equation}
This gives
\begin{equation}\label{VBRESULT22}
    v^2_{B,\mathrm{bh}} =  \frac{2\pi T_{\mathrm{bh}}}{(d-1)r_{\mathrm{bh}}}  , \quad v^2_{B,\mathrm{c}} = -\frac{2\pi T_{\mathrm{c}}}{(d-1)r_{\mathrm{c}}} .
\end{equation} 
In SdS, the presence of two horizons therefore naturally leads to two distinct Lyapunov exponents and butterfly velocities. We note that \eqref{VBRESULT2} is consistent with the result obtained for dS spacetime~\cite{Aalsma:2020aib, Grozdanov:2023txs}, where $\lambda_L = 2\pi T_\mathrm{c}$.\footnote{See \cite{Giataganas:2021ghs} for the alternative approach to Lyapunov exponent, the chaotic motion of geodesics near the horizons, in the de Sitter spacetime.} More generally, this implies that the chaos bound is saturated in both cases, with each horizon’s temperature playing the relevant role. Using \eqref{SdST}, we can further express \eqref{VBRESULT22} as
\begin{equation}\label{vbfor}
    v_{B,\mathrm{bh}}^2  L^2 = \frac{d}{2(d-1)}\frac{r_{\text{cr}}^2-r_\mathrm{bh}^2}{r_\mathrm{bh}^2}, \quad 
    v_{B,\mathrm{c}}^2  L^2 =-\frac{d}{2(d-1)}\frac{r_{\mathrm{c}}^2-r_{\text{cr}}^2}{r_{\mathrm{c}}^2} .
\end{equation}
These expressions depend on the black hole mass $M$. For instance, in the case of $d=3$, applying \eqref{rhrcRELATION}, we can expand \eqref{vbfor} in the small-mass limit as
\begin{align}\label{vbfor2}
v_{B, \mathrm{bh}}^2  L^2 = \frac{1}{16} \frac{L^2}{M^2} + \mathcal{O}\left({M}/{L}\right) , \quad 
v_{B, \mathrm{c}}^2  L^2 = -\frac{1}{2} + \mathcal{O}\left({M}/{L}\right) .
\end{align}
The behavior of these butterfly velocities is illustrated in Figure~\ref{fig:masshorizonvB}, where the small-$M$ limit is consistent with the above expansion.

Several interesting observations follow from these `hypothetical' butterfly velocities in SdS black holes, Eq.~\eqref{vbfor}, as identified by the chaotic pole-skipping points \eqref{VBRESULT}:
\paragraph{Mass dependence.}
Unlike in the planar Schwarzschild-AdS black hole case, where the butterfly velocity is independent of the mass and
\begin{equation}
 (\text{AdS result}):\quad v_B^2 L^2 =\frac{d}{2(d-1)} ,
\end{equation}
in SdS, the butterfly velocities depend on the black hole mass.\footnote{Mass dependence in the butterfly velocity has also been observed for hyperbolic Schwarzschild-AdS black holes~\cite{Ahn:2019rnq}.}
\paragraph{Imaginary cosmological butterfly velocity.}
The butterfly velocity associated with the cosmological horizon ($v_{B, \mathrm{c}} L$) is imaginary in general, i.e.,  $v_{B, \mathrm{c}}^2 L^2 \leq0$. This suggests that the OTOCs exhibit spatial oscillations. Such behavior may be related to the time oscillations previously observed in~\cite{Aalsma:2020aib} in the presence of the cosmological horizon, here, manifesting as spatial oscillations even \emph{before} the scrambling time.
\paragraph{Superluminal chaotic propagation for small black holes.}
The butterfly velocity identified at the black hole horizon  ($v_{B,\mathrm{bh}} L$) can exceed the speed of light as the black hole mass decreases. This suggests that any hypothetical holographic dual should presumably become highly nonlocal in this regime.

\vspace{6pt}
We will return to these points in Sec.~\ref{sec:interp}, where we attempt to provide a concrete microscopic interpretation of our results.

%
\subsection{Shock wave analysis}\label{sec43}
Next, we perform a shock wave analysis for SdS black holes and check whether it aligns with the pole-skipping results. Although the shock wave formalism is well established in the context of AdS black holes, particularly for extracting the Lyapunov exponent $\lambda_L$ and the butterfly velocity $v_B$~\cite{Shenker:2013pqa,Shenker:2013yza,Roberts:2014isa,Shenker:2014cwa,Grozdanov:2018kkt}, it has also been explored in other gravitational settings. In asymptotically flat spacetimes, early studies include \cite{Aichelburg1971,DRAY1985173}. The extension to spacetimes with a positive cosmological constant was first investigated by Hotta and Tanaka \cite{Hotta_1993,Hotta:1993aa} and by Sfetsos \cite{Sfetsos:1994xa}.\footnote{More recently, shock wave analysis in SdS black holes has been studied in various contexts, such as traversable wormholes \cite{Geng:2020kxh} and holographic complexity \cite{Baiguera:2023tpt,Aguilar-Gutierrez:2023pnn}. In the latter, the authors examined a shock wave intersecting the stretched horizon at a finite time. Our focus here, by contrast, is on shock waves sent at early times. Our approach can be framed in the context of a holographic description of the degrees of freedom associated with the stretched horizon, often referred to as static patch holography \cite{Susskind:2021esx,Shaghoulian:2021cef,Shaghoulian:2022fop,Dyson:2002pf,Susskind:2021omt,Susskind:2021dfc,Susskind:2011ap}.}
In this section, we follow the approach of \cite{Sfetsos:1994xa} and discuss $\lambda_L$ and $v_B$ in the context of SdS black holes.

\paragraph{Kruskal coordinates.}
For the shock wave analysis, it is useful to introduce Kruskal coordinates. Since the SdS black holes contain both a black hole horizon $r_\mathrm{bh}$ and a cosmological horizon $r_\mathrm{c}$, it is convenient to define two separate pairs of Kruskal coordinates: $(U,V)=(U_\mathrm{bh},V_\mathrm{bh})$ for $r_\mathrm{bh}$ and $(U_\mathrm{c},V_\mathrm{c})$ for $r_\mathrm{c}$. For instance, in $d=3$~\cite{Bhattacharya:2018ltm}, these are related to the Eddington-Finkelstein coordinates $(u,v)$ as follows:
\begin{align}\label{KCeq0}
\begin{split}
(U_\mathrm{bh}, V_\mathrm{bh}) = \frac{1}{\kappa_\mathrm{bh}}\left(-e^{-\kappa_\mathrm{bh} u}, e^{\kappa_\mathrm{bh} v}\right),
\quad
(U_\mathrm{c}, V_\mathrm{c}) = \frac{1}{\kappa_\mathrm{c}}\left(e^{\kappa_\mathrm{c} u}, -e^{-\kappa_\mathrm{c} v}\right),
\end{split}
\end{align}
where $\kappa_\mathrm{bh}$ and $\kappa_\mathrm{c}$ denote the surface gravities at the black hole and cosmological horizons, respectively,
\begin{equation}
\kappa_\mathrm{bh} = \Lambda \frac{(2r_\mathrm{bh} + r_\mathrm{c})(r_\mathrm{c}-r_\mathrm{bh})}{6r_\mathrm{bh}}, \quad 
\kappa_\mathrm{c} = \Lambda\frac{(2r_\mathrm{c} + r_\mathrm{bh})(r_\mathrm{c}-r_\mathrm{bh})}{6r_\mathrm{c}},
\end{equation}
where $\Lambda$ is the cosmological constant given in \eqref{Action:EH}.
These Kruskal coordinates \eqref{KCeq0} satisfy the relations
\begin{equation}\label{UVUVRE}
\begin{split}
        U_\mathrm{bh} V_\mathrm{bh} = -\frac{1}{\kappa_\mathrm{bh}^2}e^{\kappa_\mathrm{bh}(v-u)}=-\frac{1}{\kappa_\mathrm{bh}^2}e^{2\kappa_\mathrm{bh} r_*},\quad
        U_\mathrm{c} V_\mathrm{c} = -\frac{1}{\kappa_\mathrm{c}^2}e^{\kappa_\mathrm{c}(u-v)}=-\frac{1}{\kappa_\mathrm{c}^2}e^{-2\kappa_\mathrm{c} r_*},
\end{split}
\end{equation}
where $r_*$ is the tortoise coordinate. 

\begin{figure}[]
  \centering
     {\includegraphics[width=9.4cm]{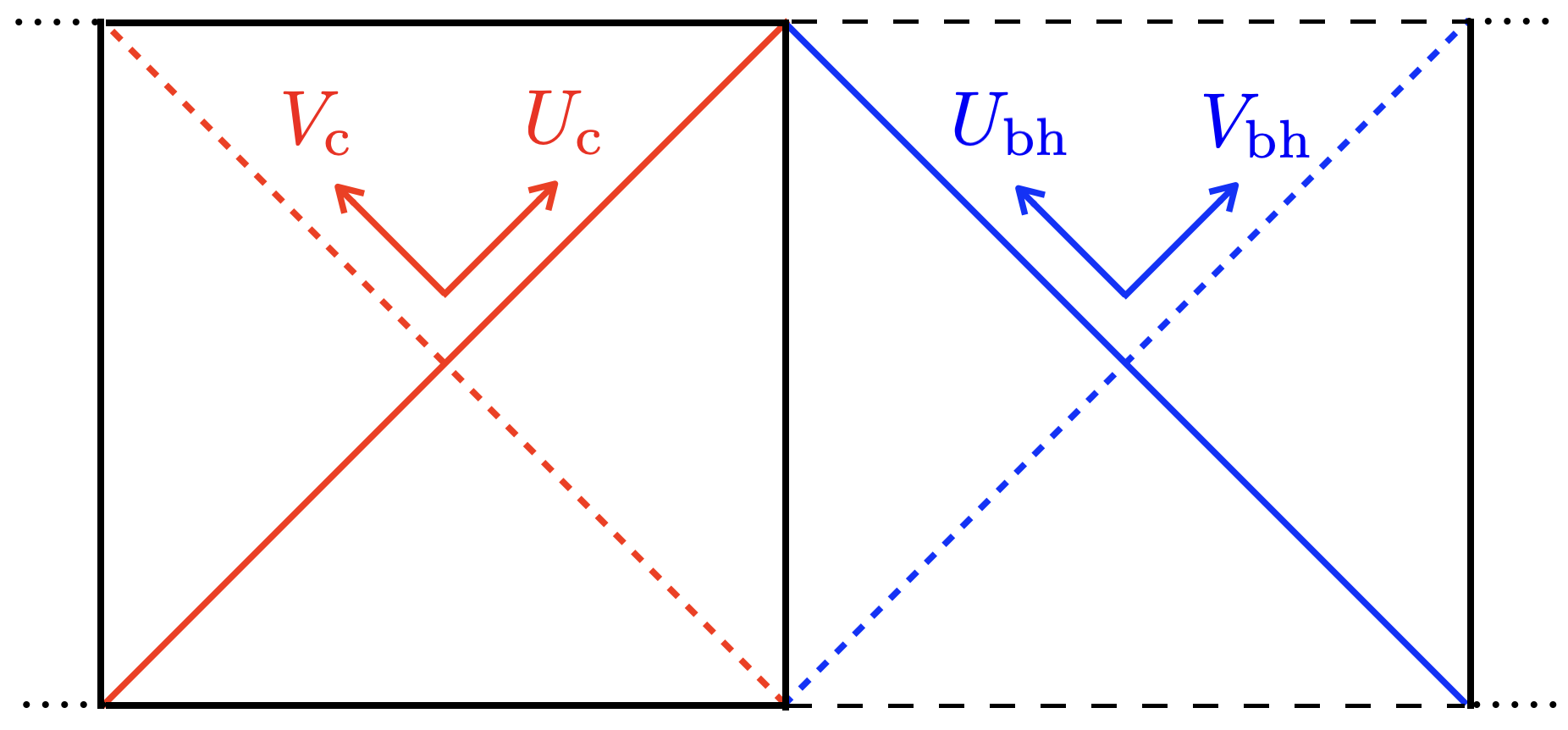} \label{}}
 \caption{Penrose diagram of SdS$_{d+1}$ black holes when $d\geq3$ in Kruskal coordinates: \eqref{KCeq1} and \eqref{KCeq2}. The shock waves at $U_{\mathrm{bh}|\mathrm{c}}=0$ are represented as the blue and red dashed lines.}\label{PENROSEFIG3}
\end{figure}

In terms of these Kruskal coordinates, our metric \eqref{Ansatz:INEF} with \eqref{Sol:f} takes the form
\begin{align}\label{KCeq1}
\begin{split}
    \dd s^2 &=-\frac{2M}{r}\left(1-\frac{r}{r_\mathrm{c}}\right)^{1+\frac{\kappa_\mathrm{bh}}{\kappa_\mathrm{c}}}\left(1+\frac{r}{r_\mathrm{bh}+r_\mathrm{c}}\right)^{1+\frac{\kappa_\mathrm{c}}{\kappa_u}}\dd U_\mathrm{bh} \dd V_\mathrm{bh} + r^2   \gamma_{ij}\dd x^i \dd x^j ,
\end{split}
\end{align}
for $(U,V)=(U_\mathrm{bh},V_\mathrm{bh})$, and
\begin{align}\label{KCeq2}
\begin{split}
    \dd s^2 &=-\frac{2M}{r}\left(\frac{r}{r_\mathrm{bh}}-1\right)^{1+\frac{\kappa_\mathrm{c}}{\kappa_\mathrm{bh}}}\left(1+\frac{r}{r_\mathrm{bh}+r_\mathrm{c}}\right)^{1+\frac{\kappa_\mathrm{c}}{\kappa_u}}\dd U_\mathrm{c} \dd V_\mathrm{c} + r^2   \gamma_{ij}\dd x^i \dd x^j ,
\end{split}
\end{align}
for $(U_\mathrm{c},V_\mathrm{c})$. Here, $\kappa_u=(M/r_u^2-\Lambda r_u/3)$ is the surface gravity associated with the unphysical negative horizon $r_u=-(r_\mathrm{bh}+r_\mathrm{c})$. See Figure~\ref{PENROSEFIG3} for the corresponding Penrose diagram.
For simplicity, hereafter, we express the Kruskal coordinates \eqref{KCeq1}--\eqref{KCeq2} in the following unified form:
\begin{align}\label{KC1}
\begin{split}
\dd s^2 = 2 A(U, V)  \dd U \dd V + r^2(U,V)  \gamma_{ij}\dd x^i \dd x^j .
\end{split}
\end{align}
This coordinate system can also be derived from \eqref{Ansatz:INEF} in other dimensions, where the function $A(U,V) = \frac{\beta^2}{8\pi^2}\frac{f(U,V)}{U V}$ with the relations $UV = -e^{\frac{4\pi}{\beta} r_{*}}$ and $U/V = -e^{-\frac{4\pi}{\beta} t}$, where $\beta$ denotes the inverse temperature.

\paragraph{Shock wave geometry.}
To analyze the effect of a shock wave on the metric \eqref{KC1}, we introduce a massless particle moving at the speed of light along the $V$-direction at $U=0$, as indicated by the red and blue dashed lines in Figure~\ref{PENROSEFIG3}.

Our objective is to determine the impact of the shock wave on the background geometry \eqref{KC1}. Specifically, we assume that for $U<0$, the spacetime is described by \eqref{KC1}, while for $U>0$, the metric remains the same except for a shift in $V$, given by $V\rightarrow V+h(x^i)$, where $h(x^i)$ is a function to be determined.
As a result, the modified spacetime metric takes the form
\begin{align}\label{METAFSC}
\begin{split}
\dd s^2 &= 2 A\left(U, V + \Theta(U) h(x^{i}) \right) \dd U \left[\dd V + \Theta(U) \dd h \right] + r^2\left(U, V + \Theta(U) h(x^{i}) \right)  \gamma_{ij}\dd x^i \dd x^j ,
\end{split}
\end{align}
where $\Theta(U)$ is the Heaviside step function. 

\paragraph{Einstein's equations at first order.} The shock wave at $U=0$ can be generated by a null energy pulse with the energy-momentum tensor
\begin{align}\label{shockTUU}
\begin{split}
T^{\text{Shock}}_{UU} = 4 A^2  p  \,\delta(U)  \delta^{d-2}(x^i) ,
\end{split}
\end{align}
where $p$ represents the massless particle's momentum.\footnote{For a null particle propagating in the background \eqref{METAFSC}, the only relevant nonvanishing component of the stress tensor is $T_{UU}$, which follows from
$
T_{UU}=g_{U\mathcal V}\,T^{\mathcal V\mathcal V}\,g_{\mathcal VU}=4A^{2}p.
$
In the eikonal regime, an early-time perturbation is exponentially localized near the horizon in Kruskal coordinates, so it is natural to idealize its stress tensor as a null delta-function source at $U=0$. Allowing the source to be supported instead at $U=U_{0}$, with $|U_{0}|\ll 1$, changes the result only through subleading finite-boost corrections and does not modify the leading values of $\lambda_{L}$ or $v_{B}$~\cite{Shenker:2013pqa,Shenker:2013yza,Shenker:2014cwa,Dray:1984ha,tHooft:1987vrq,Cornalba:2006xk} in AdS backgrounds.} Here, for a particle released at an early time $t$ and falling into the horizon, its momentum undergoes an exponential boost
\begin{equation}\label{ptime}
p \propto e^{2\pi T_{\mathrm{bh}|\mathrm{c} } t} .
\end{equation}

To determine the backreaction of the shock wave, we solve Einstein's equation \eqref{Eq:EE} for the perturbed metric \eqref{METAFSC} with the energy-momentum tensor
\begin{align}\label{METAFSC2}
\begin{split}
T_{UU}^{(1)} =   {T}^{\text{Shock}}_{UU}   .
\end{split}
\end{align}
This leads to the following equation:
\begin{align}\label{FRE1}
\begin{split}
\left(\hat{D}_i \hat{D}^{i} - \frac{d-1}{2} \frac{\partial_{UV} r^2}{A} \right) h(x^i) = 32 \pi G_N  A  p  r^2    \delta^{d-2}(x^i)  ,
\end{split}
\end{align}
which is evaluated at $U=0$ or $r=r_{\mathrm{bh}|\mathrm{c}}$. Using the definition of the Kruskal and tortoise coordinates, one can find the relation
\begin{equation}
    \left.\partial_{UV}r^2\right|_{r=r_{\mathrm{bh}|\mathrm{c}}}=\left. A f'(r) r\right|_{r=r_{\mathrm{bh}|\mathrm{c}}} = \pm  4\pi A_{\mathrm{bh}|\mathrm{c}} T_{\mathrm{bh}|\mathrm{c}} r_{\mathrm{bh}|\mathrm{c}} .
\end{equation}
Substituting this into \eqref{FRE1}, we obtain 
\begin{equation}\label{Eq: shock wave}
    \left(\hat{D}_i\hat{D}^i \mp m^2 \right)h(x^i) = 32\pi G_N  A_{\mathrm{bh}|\mathrm{c}} p r_{\mathrm{bh}|\mathrm{c}}^2  \delta^{(d-2)}(x^i) ,
\end{equation}
where the negative sign corresponds to $r_\mathrm{bh}$, while the positive sign corresponds to $r_\mathrm{c}$, with
\begin{align}\label{mTrTc}
\begin{split}
m^2 =  2\pi\left(d-1\right) T_{\mathrm{bh}|\mathrm{c}} r_{\mathrm{bh}|\mathrm{c}}  ,
\end{split}
\end{align}
The equation \eqref{Eq: shock wave} characterizes the effect of the shock wave on the spacetime geometry.

\paragraph{Butterfly velocities from the shock wave profile.}
We now examine the relationship between OTOCs and the shock wave profile $h(x^{i})$ in \eqref{Eq: shock wave}, aiming to extract the Lyapunov exponent $\lambda_L$ and the butterfly velocity $v_B$. We note that the problem exhibits spherical symmetry. To simplify the analysis, we therefore consider a source located at a point on the equator of the transverse $\mathcal{S}^{d-1}$ and hold all angles fixed except for the azimuthal angle. In the coordinates of (\ref{Ansatz:INEF}) this corresponds to setting $\chi=1$, $\theta_i=\pi/2$ ($i\neq1$) and letting $h$ depend on $\theta_1 \in [0,2\pi]$. Equation~\eqref{Eq: shock wave} thus reduces away from the source to
\begin{equation}\label{Eq:shockwave2}
    h''(\theta_1) \mp m^2 h(\theta_1) = 0 ,
\end{equation}
which can be solved analytically.

\begin{figure}[]
  \centering
     {\includegraphics[width=7.2cm]{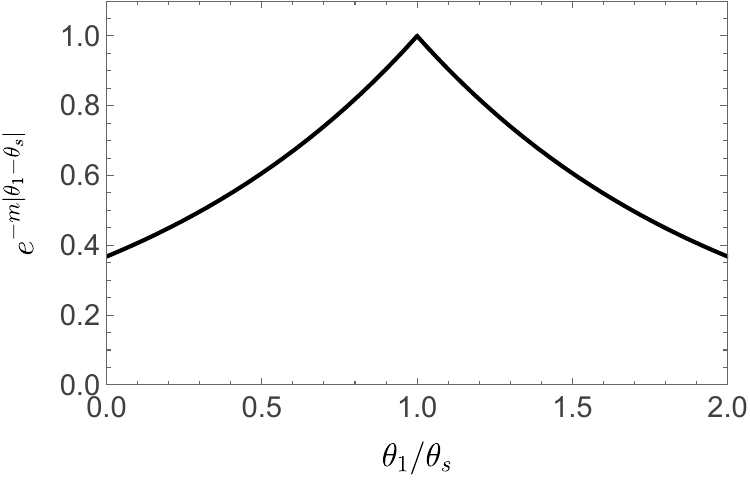} \label{}}
\quad
     {\includegraphics[width=7.2cm]{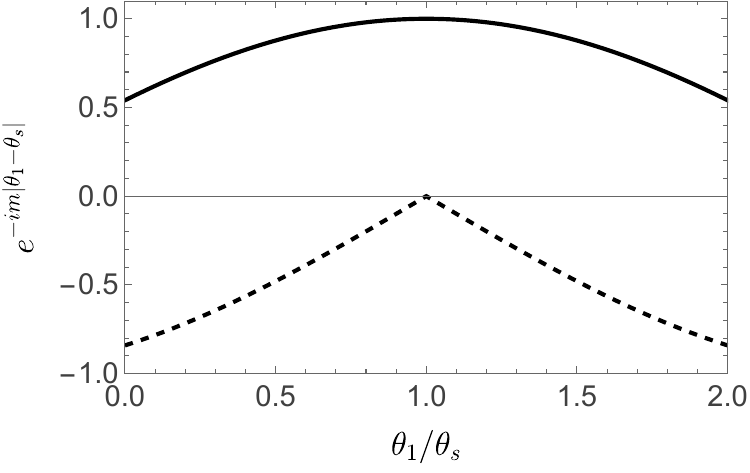} \label{}}
 \caption{Schematic representation of the spatial profiles of the shock wave solution in \eqref{arg11} (left) and \eqref{arg22} (right) for the parameter values $m =1/\pi$, and $\theta_s=\pi$. In the right panel, the solid line corresponds to the real part of the solution, while the dashed line represents the imaginary part.}\label{SWFIG}
\end{figure}

\paragraph{\quad (I) Black hole horizon.}
For the black hole horizon (corresponding to the negative sign in \eqref{Eq:shockwave2}), the general solution takes the form
\begin{align}
\begin{split}
h(t, \theta_1) 
&\approx \exp \left( 2\pi T_{\mathrm{bh}} t \right)
\left[
c_1 \exp\left( m \theta_1 \right)
+
c_2 \exp\left(- m \theta_1 \right)
\right].
\end{split}
\end{align}
As usual, the time dependence follows from the boost factor \eqref{ptime}, while the coefficients \(c_i\) are undetermined constants. Assuming further that the perturbation originates from a localized source at \(\theta_1=\theta_s\), the solution depends on the sign of \(\theta_1-\theta_s\). One then finds
\begin{align}\label{arg11}
\begin{split}
h(t, \theta_1)
\sim 
\exp \left( 2\pi T_{\mathrm{bh}} t \right)
\exp\left(- m  \left|\theta_1-\theta_s\right| \right),
\end{split}
\end{align}
i.e., the solution decays exponentially away from \(\theta_s\), as illustrated in the left panel of Figure~\ref{SWFIG}.
We may compare \eqref{arg11} with the expected behavior from an OTOC, namely
\begin{align}\label{OTOCTX2}
\begin{split}
\text{exp} \left( \lambda_L  t \right) \text{exp} \left( -\frac{\lambda_L}{v_B}  |x| \right).
\end{split}
\end{align}
Identifying $\left|\theta_1-\theta_s\right|$ with $|x|$,\footnote{We note that for spherical horizons \eqref{VBRESULT} naturally defines an angular velocity. If we want a linear velocity, we can simply multiply $v_B$ by the curvature radius $L$, as in \eqref{vbfor}.} we obtain
\begin{align}\label{VBSWBH}
\begin{split}
\lambda_{L} = 2\pi T_{\mathrm{bh}} , \quad  v^2_{B,\mathrm{bh}} =  \frac{\lambda_{L}^2}{m^2} =  \frac{2\pi T_{\mathrm{bh}}}{\left(d-1\right) r_{\mathrm{bh}}} ,
\end{split}
\end{align}
where we have used Eq.~\eqref{mTrTc}. This result is consistent with the pole-skipping analysis at the black hole horizon, given in Eqs.~\eqref{VBRESULT2} and \eqref{VBRESULT22}.

\paragraph{\quad (II) Cosmological horizon.}
For the cosmological horizon (corresponding to the positive sign in \eqref{Eq:shockwave2}), the solution takes a different form:
\begin{align}\label{arg22}
\begin{split}
h(t, \theta_1) 
&\approx \exp \left( 2\pi T_{\mathrm{c}} t \right)
\left[
c_1 \cos\left( m \theta_1 \right)
+
c_2 \sin\left(m \theta_1 \right)
\right],\\
&\sim \exp \left( 2\pi T_{\mathrm{c}} t \right)\exp\left( \pm i m |\theta_1-\theta_s| \right) .
\end{split}
\end{align}
In the second line, we express the solution in terms of plane waves, using an argument similar to that leading to \eqref{arg11}. The right panel of Figure~\ref{SWFIG} provides a schematic illustration.

Unlike the black hole case, this solution does not match the OTOC form \eqref{OTOCTX2}. However, if one formally replaces the spatial decay term according to \(v_B \rightarrow \pm i \tilde{v}_B\), one obtains
\begin{align}
\begin{split}
\exp \left( \lambda_L t \right)\exp \left( \pm i \frac{\lambda_L}{\tilde{v}_B} |x| \right),
\end{split}
\end{align}
which is consistent with \eqref{arg22}. This leads to
\begin{align}\label{VBCSW}
\begin{split}
\lambda_{L} = 2\pi T_{\mathrm{c}}, \quad
v^2_{B,\mathrm{c}} = -\tilde{v}^2_{B,\mathrm{c}} = -\frac{\lambda_{L}^2}{m^2}
= -\frac{2\pi T_{\mathrm{c}}}{\left(d-1\right) r_{\mathrm{c}}}.
\end{split}
\end{align}
This shock wave result is again consistent with the pole-skipping analysis at the cosmological horizon \(r_\mathrm{c}\), as follows from Eqs.~\eqref{VBRESULT22} and \eqref{VBRESULT2}.

\vspace{6pt}

In summary, both pole-skipping and shock wave analyses yield consistent expressions for $\lambda_L$ and $v_B$ at the black hole and cosmological horizons. Notably, our discussion includes the aforementioned superluminal butterfly velocity \eqref{VBSWBH} for small black holes and the imaginary butterfly velocity \eqref{VBCSW} for the cosmological horizon, both found in Section~\ref{sec42} and depicted in Figure~\ref{fig:masshorizonvB}. In the next section, we explore potential microscopic interpretations of these results.

%
\subsection{Toward a microscopic holographic interpretation}\label{sec:interp}

\subsubsection{Holographic screens}

In the search for a holographic description of the SdS spacetime, a helpful notion is that of a \textit{holographic screen}: a surface on which bulk degrees of freedom can be encoded~\cite{Bousso:1999cb}.\footnote{For an early proposal of holographic screens in SdS, see~\cite{Chang-Young:2010lou}. The term `screen', which is a codimension-one surface where the higher-dimensional physical system is `imprinted', is closely analogous to the concept of a screen in the theory of film and photography, see e.g.,~\cite{huhtamo2004elements,db271f62-4a9d-3530-822a-0fca0cffc7d0,polonyi2024opportunity,Polonyi2019}.
}

In standard AdS/CFT holography, the conformal boundary at spatial infinity plays this role: the entire bulk is dual to a CFT defined on that screen. When the geometry contains horizons, one may extend the idea to treat each horizon as an additional screen. To see how, it is helpful to start with a purification of the system. For an eternal AdS black hole, for instance, the Penrose diagram has two disconnected asymptotic boundaries, and the dual description involves two entangled copies of the same CFT in a Thermofield Double (TFD) state~\cite{Maldacena:2001kr}. If we restrict attention to one exterior region only, namely we `integrate out' part of the geometry~\cite{Faulkner:2010jy,Guijosa:2022jdo}, the would‑be degrees of freedom of the second CFT would then be encoded on the black hole horizon, which acts as an effective holographic screen. The price we pay for integrating out one side is that the degrees of freedom of the second CFT are now encoded non‑locally on the horizon.

By analogy, SdS spacetime naturally contains two such screens: the black hole event horizon and the cosmological horizon. These horizons bound the static patch accessible to an observer. Assigning an entropy $S_{\text{bh}|\text{c}}=A_{\text{bh}|\text{c}}/4G_N$ to each horizon, one may regard them as two quantum mechanical systems whose combined state is entangled, mirroring the TFD picture of the eternal AdS black hole. Individually, each system appears thermal, with the temperature $T_i=|\kappa_i|/2\pi$ set by its own surface gravity, but because $T_{\mathrm{bh}}\neq T_{\mathrm{c}}$ generically, no global equilibrium exists except in the extremal Nariai limit in which the two horizons coincide. 

Accordingly, we may posit an approximate Hilbert‑space factorization at the semi-classical level,
$\mathcal{H}_{\text{total}}\simeq\mathcal{H}_{\mathrm{bh}}\otimes\mathcal{H}_{\mathrm{c}}$,
with Hamiltonians $H_{\mathrm{bh}}$ and $H_{\mathrm{c}}$, prepared in an arbitrary entangled state and interactions suppressed at leading order in $G_N$ (see also \cite{Begines:2026fnx}),\footnote{In this picture, the black hole and cosmological horizons may be viewed as defining two effective horizon-associated sectors. A more careful formulation may require doubling the degrees of freedom,
$\mathcal{H}_{\text{total}}\simeq\mathcal{H}_{\mathrm{bh}}\otimes\mathcal{H}_{\mathrm{bh'}}\otimes\mathcal{H}_{\mathrm{c}}\otimes\mathcal{H}_{\mathrm{c'}}$ so that thermal or mixed states arise by tracing out complementary sectors. In pure de Sitter this leaves only the cosmological sector (and its complement), while in the Nariai limit the equal horizon temperatures make the state more TFD-like. Unlike the eternal AdS black hole, however, these screens are not asymptotic boundaries, and the construction should therefore be regarded as an effective organizing picture rather than a complete microscopic duality.}
\begin{equation}
  \ket{\Psi}
  =\sum_{n,m} c_{nm}
  \ket{E_{n}^{(\mathrm{bh})}}\!\otimes\!\ket{E_{m}^{(\mathrm{c})}} .
\end{equation}
Unlike in the TFD, this state need not be symmetric, because there is no direct relation between the spectra of $H_{\mathrm{bh}}$ and $H_{\mathrm{c}}$. Tracing out, say, $\mathcal{H}_{\mathrm{c}}$, gives
\begin{equation}
  \rho_{\mathrm{bh}}
  = \operatorname{Tr}_{\mathrm{c}}\!\bigl[\ket{\Psi}\bra{\Psi}\bigr],
\end{equation}
which may look thermal with respect to $H_{\mathrm{bh}}$, yet generically with a different temperature from that obtained by tracing out the opposite side. The global state remains pure, but the reduced states live at distinct effective temperatures, reflecting the unequal surface gravities of the two horizons.

\paragraph{Example:} As a simple illustration using qubit systems, consider the following toy model:
\begin{align}
H_{i} &= \Delta_{i}
|1\rangle\langle 1| \quad i=\{\text{bh, c}\}.
\end{align}
Both Hamiltonians assign energy 0 to $|0\rangle$ and a gap $\Delta_i$ to $|1\rangle$, but with distinct gaps,
\begin{equation}
\Delta_{\text{bh}}=1,\quad \Delta_{\text{c}}=3.
\end{equation}
Now, consider the Schmidt‑rank‑2 state
\begin{equation}
\ket{\Psi} = \sqrt{p}\ket{0,0} + \sqrt{1-p}\ket{1,1},
\quad p=\tfrac{2}{3}.
\end{equation}
Because the amplitudes are correlated, tracing out either qubit gives a \emph{mixed} state, but the populations are the same on both sides. Tracing out the subsystem ``c'' yields
\begin{equation}
\rho_{\mathrm{bh}} 
  = p|0\rangle\langle 0| + (1-p)|1\rangle\langle 1|.
\end{equation}
The excited‑to‑ground population ratio is $r=(1-p)/p=1/2$.  For a two-level system with the Hamiltonian $H=\Delta|1\rangle\langle1|$, a Gibbs state at inverse temperature $\beta$ has $r = e^{-\beta\Delta}$.  Hence,  
\begin{equation}
\beta_{\mathrm{bh}} = \frac{\ln 2}{\Delta_{\mathrm{bh}}}
                    = \ln 2 \approx 0.693 ,\quad
T_{\mathrm{bh}} \equiv \beta_{\mathrm{bh}}^{-1}\approx 1.44 .
\end{equation}
Conversely, tracing out the subsystem ``bh'' yields an identical density matrix, $
\rho_{\mathrm{c}} = \rho_{\mathrm{bh}}$, but its Hamiltonian has a \emph{larger} gap $\Delta_{\mathrm{c}}=3$.   The effective temperature therefore differs:
\begin{equation}
\beta_{\mathrm{c}} = \frac{\ln 2}{\Delta_{\mathrm{c}}}
                   = \frac{\ln 2}{3}\approx 0.231 ,\quad
T_{\mathrm{c}}\approx 4.33 .
\end{equation}

\vspace{8pt}

In practice, however, the quantum mechanical models dual to the two holographic screens are expected to be large-$N$ systems. We know little about the theories dual to dS spacetimes, but the bulk calculations of the Lyapunov exponent and the butterfly velocity allow us to infer some of their peculiar defining properties. Below, inspired by our results from the earlier sections, we will propose concrete models that could capture some of these properties.

\subsubsection{DSSYK chain with long-range interactions and the superluminal butterfly velocity}

Motivated by the pole-skipping and shock-wave analyses of Sections \ref{sec42} and \ref{sec43}, respectively, we now develop a microscopic toy model capable of reproducing the enhancement of the butterfly velocity $v_{B,\text{bh}}$ as the black-hole horizon shrinks, together with the emergence of a superluminal chaotic spread. Our starting point is the Sachdev--Ye--Kitaev (SYK) model of $N$ real Majorana fermions interacting through all-to-all $q$-body random interactions. The so-called double-scaled limit (DSSYK), where $N\to\infty$ and $q\to\infty$ with $q^2/N$ fixed, is conjectured to be dual to dS$_2$~\cite{Susskind:2022dfz,Lin:2022nss,Lin:2022rbf,Susskind:2022bia,Rahman:2022jsf,Narovlansky:2023lfz,Verlinde:2024znh,Aguilar-Gutierrez:2025mxf}. In what follows we restrict attention to the IR conformal window in which the four-point function is governed by the same effective retarded-kernel logic as in the ordinary large-$N$ SYK$_q$ chain, even though the exact DSSYK four-point function is more naturally organized in terms of chord/skeleton diagrams. For our purposes, only this IR chaotic data will be needed.\footnote{We use ``IR CFT'' in the operational sense, since DSSYK exhibits an emergent IR conformal regime of correlators; for a UV-complete avatar, one may instead use the triple-scaled limit~\cite{Rabinovici:2023yex,Ambrosini:2024sre}. Since our discussion only relies on IR chaotic data, the ladder analysis below should be understood as the corresponding low-energy effective description.}

\begin{figure}[t]
\centering
\begin{tikzpicture}[scale=1, every node/.style={font=\small}]
  \foreach \x in {0,2,4,6,8} {
    \draw[rounded corners] (\x,0) rectangle (\x+1.3,0.9);
    \node at (\x+0.65,0.45) {$H_{\rm SYK}$};
    \fill (\x+0.15,0.15) circle (1.2pt);
    \fill (\x+1.15,0.15) circle (1.2pt);
    \fill (\x+0.15,0.75) circle (1.2pt);
    \fill (\x+1.15,0.75) circle (1.2pt);
  }
  \foreach \x in {1.3,3.3,5.3,7.3} {
    \draw[-,thick] (\x,0.45) -- (\x+0.7,0.45);
  }
  \draw[thick,dashed] (1.3,0.8) to[bend left=25] (6,0.8);
  \node at (3.7,1.65) {$H_{\rm hop}$};
\end{tikzpicture}
\caption{Schematic chain of SYK sites coupled by an inter-site hopping term $H_{\rm hop}$.}
\label{fig:syk_chain_cartoon}
\end{figure}

To emulate a higher-dimensional bulk, we assemble a chain of $L$ identical sites labeled $j=1,\dots,L$. See Fig.~\ref{fig:syk_chain_cartoon} for a schematic representation. Each site carries $N$ Majoranas $\psi_{j,a}$ ($a=1,\dots,N$) with an on-site Hamiltonian identical to that of the SYK model:
\begin{equation}\label{eq:DSSYK}
  H^{(j)}_{\text{SYK}}
     = i^{q/2}
       \sum_{1\le a_{1}<\cdots<a_{q}\le N}
       J^{(j)}_{a_{1}\cdots a_{q}}
       \psi_{j,a_{1}}\!\dots\psi_{j,a_{q}},
\end{equation}
where the random couplings are Gaussian-distributed with variance
\begin{equation}
\langle
       (J_{a_1\cdots a_q}^{(j)})^2
   \rangle
   =
   \frac{(q-1)!J_0^{2}}{N^{q-1}}.
\end{equation}
Here, $J_0$ sets the overall energy scale of on-site interactions. Spatial non-locality is now introduced through random inter-site couplings that decay with the distance between sites,
\begin{equation}\label{eq:Hhop}
H_{\text{hop}}=i\sum_{j<k}\sum_{a=1}^{N}\frac{I^{(a)}_{jk}}{|j-k|^{\alpha}}\psi_{j,a}\psi_{k,a},
\end{equation}
with
\begin{equation}\label{eq:HhopVar}
\langle (I^{(a)}_{jk})^2\rangle=J_{1}^{2}.
\end{equation}
In other words, a Majorana fermion at site $j$ can hop to site $k$ with amplitude that falls off as a power law $1/|j-k|^{\alpha}$, where $J_1$ controls the overall strength of the hopping. The exponent $\alpha>0$ tunes the interaction range: $\alpha\gg1$ corresponds to effectively nearest-neighbor hoppings, whereas $\alpha\to0$ approaches all-to-all couplings along the chain.

The total Hamiltonian then reads
\begin{equation}
H=\sum_{j=1}^L H^{(j)}_{\text{SYK}} + H_{\text{hop}}.
\end{equation}
At large $N$, the on-site chaotic dynamics is unaltered by $H_{\text{hop}}$ at leading order in $1/N$: each site remains maximally chaotic with Lyapunov exponent $\lambda_{L}=2\pi/\beta$. In other words, adding the hopping (if appropriately scaled with $N$) does not modify the local scrambling rate at leading order, because the local on-site dynamics continues to control the temporal growth. What does change, however, is the \emph{spatial} propagation of chaos through the chain.

\begin{figure}[t]
\centering
\begin{tikzpicture}[x=1cm,y=1cm,baseline={(current bounding box.center)}]

\node at (-1.2,0.7) {$\mathcal F \,=$};

\begin{scope}[shift={(0,0)}]
  \draw[thick] (0,1.2) -- (1.8,1.2);
  \draw[thick] (0,0.2) -- (1.8,0.2);
  \fill (0,1.2) circle (1.2pt) node[left] {$\tau_1$};
  \fill (0,0.2) circle (1.2pt) node[left] {$\tau_2$};
  \fill (1.8,1.2) circle (1.2pt) node[right] {$\tau_3$};
  \fill (1.8,0.2) circle (1.2pt) node[right] {$\tau_4$};
  \node at (0.9,-0.35) {$\mathcal F_0$};
\end{scope}

\node at (2.5,0.7) {$+$};

\begin{scope}[shift={(3,0)}]
  \draw[thick] (0,1.2) -- (0.7,1.2);
  \draw[thick] (1.1,1.2) -- (1.8,1.2);
  \draw[thick] (0,0.2) -- (0.7,0.2);
  \draw[thick] (1.1,0.2) -- (1.8,0.2);
  \draw[thick] (0.9,0.2) -- (0.9,1.2);
  \fill (0,1.2) circle (1.2pt) node[left] {$\tau_1$};
  \fill (0,0.2) circle (1.2pt) node[left] {$\tau_2$};
  \fill (1.8,1.2) circle (1.2pt) node[right] {$\tau_3$};
  \fill (1.8,0.2) circle (1.2pt) node[right] {$\tau_4$};
  \node at (0.9,-0.35) {$\mathcal F_1$};
\end{scope}

\node at (5.5,0.7) {$+$};

\begin{scope}[shift={(6,0)}]
  \draw[thick] (0,1.2) -- (0.45,1.2);
  \draw[thick] (0.75,1.2) -- (1.05,1.2);
  \draw[thick] (1.35,1.2) -- (1.8,1.2);
  \draw[thick] (0,0.2) -- (0.45,0.2);
  \draw[thick] (0.75,0.2) -- (1.05,0.2);
  \draw[thick] (1.35,0.2) -- (1.8,0.2);
  \draw[thick] (0.6,0.2) -- (0.6,1.2);
  \draw[thick] (1.2,0.2) -- (1.2,1.2);
  \fill (0,1.2) circle (1.2pt) node[left] {$\tau_1$};
  \fill (0,0.2) circle (1.2pt) node[left] {$\tau_2$};
  \fill (1.8,1.2) circle (1.2pt) node[right] {$\tau_3$};
  \fill (1.8,0.2) circle (1.2pt) node[right] {$\tau_4$};
  \node at (0.9,-0.35) {$\mathcal F_2$};
\end{scope}

\node at (8.7,0.7) {$+\cdots$};

\end{tikzpicture}
\caption{First few ladder contributions to the connected four-point function. The horizontal lines are the dressed propagators (``rails''), while the vertical insertions are the interaction rungs.}
\label{fig:otoc_ladder_sum}
\end{figure}

To analyze this, we consider the connected four-point out-of-time-ordered correlator. In the large-$N$ limit it is obtained by summing an infinite series of ladder diagrams built from the dressed two-point functions. The first few contributions are shown in Fig.~\ref{fig:otoc_ladder_sum}, and we denote by $\mathcal{F}_n$ the ladder with $n$ rungs, so that
\begin{equation}
\mathcal{F}=\sum_{n=0}^{\infty}\mathcal{F}_n.
\end{equation}
Here the horizontal propagators are the ``rails'' and the vertical insertions are the ``rungs.'' The detailed form of a rung depends on $q$: for $q=4$ it is the simplest one familiar from the standard SYK chain, whereas for general SYK$_q$ it carries the usual $G^{\,q-2}$ dressing \cite{Maldacena:2016hyu}. The recursive structure is universal, however: the $(n+1)$-rung diagram is generated from the $n$-rung one by acting with the retarded ladder kernel $K_R$,
\begin{equation}
\mathcal{F}_{n+1}(\tau_1,\tau_2,\tau_3,\tau_4)
=
\int d\tau\,d\tau'\,
K_R(\tau_1,\tau_2;\tau,\tau')\,
\mathcal{F}_{n}(\tau,\tau',\tau_3,\tau_4).
\end{equation}
This is illustrated in Fig.~\ref{fig:kernel_action}. Summing the geometric series then yields the usual Bethe--Salpeter form for the Fourier transform of the four-point function,
\begin{equation}
\label{eq:ladder_resummed}
\begin{split}
f(k,\omega)
&=f^{(0)}(k,\omega)+K_R(k,\omega)f^{(0)}(k,\omega)+K^2_R(k,\omega)f^{(0)}(k,\omega)+\cdots\,\\
&=\frac{f^{(0)}(k,\omega)}{1-K_R(k,\omega)}\,,
\end{split}
\end{equation}
where $f^{(0)}(k,\omega)$ is the disconnected contribution.

\begin{figure}[t]
\centering
\begin{tikzpicture}[x=1cm,y=1cm,baseline={(current bounding box.center)}]

\begin{scope}[shift={(0,0)}]
  \draw[thick,blue] (0,1.2) -- (0.6,1.2);
  \draw[thick,blue] (0,0.2) -- (0.6,0.2);
  \draw[thick] (0.6,1.2) -- (3.0,1.2);
  \draw[thick] (0.6,0.2) -- (3.0,0.2);
  \draw[thick,blue] (0.6,0.2) -- (0.6,1.2);
  \draw[thick] (1.2,0.2) -- (1.2,1.2);
  \draw[thick] (1.8,0.2) -- (1.8,1.2);
  \draw[thick] (2.4,0.2) -- (2.4,1.2);
  \node at (1.5,-0.35) {$\mathcal F_{n+1}$};

\node at (3.7,0.7) {$=$};

\begin{scope}[shift={(4.2,0)}]
  \draw[thick,blue] (0,1.2) -- (0.5,1.2);
  \draw[thick,blue] (0,0.2) -- (0.5,0.2);
  \draw[thick,blue] (0.5,0.2) -- (0.5,1.2);
  \node[blue] at (0.5,-0.35) {$K_R$};
\end{scope}

\node at (5.3,0.7) {$\circ$};

\begin{scope}[shift={(5.8,0)}]
  \draw[thick] (0,1.2) -- (2.4,1.2);
  \draw[thick] (0,0.2) -- (2.4,0.2);
  \draw[thick] (0.6,0.2) -- (0.6,1.2);
  \draw[thick] (1.2,0.2) -- (1.2,1.2);
  \draw[thick] (1.8,0.2) -- (1.8,1.2);
  \node at (1.2,-0.35) {$\mathcal F_n$};
\end{scope}

\end{scope}

\end{tikzpicture}
\caption{Recursive generation of ladder diagrams by the retarded kernel. Acting with $K_R$ adds one extra rung, $\mathcal F_{n+1}=K_R\circ\mathcal F_n$.}
\label{fig:kernel_action}
\end{figure}

Next, it is convenient to Fourier transform the coordinate along the chain,
\(
\psi_{k,a}\equiv L^{-1/2}\sum_{j}e^{-ikj}\psi_{j,a}
\)
(with \(k\in\tfrac{2\pi}{L}\mathbb{Z}\)). The large-\(N\) Schwinger--Dyson equations for the site-diagonal two-point function,
\(
G(k,\tau)=\tfrac{1}{N}\sum_a\langle \mathcal{T}\psi_{k,a}(\tau)\psi_{-k,a}(0)\rangle,
\)
take the standard form
\begin{equation}
G^{-1}(k,i\omega_n)=i\omega_n-\Sigma_{\rm SYK}(i\omega_n)-\Sigma_{\rm hop}(k,i\omega_n).
\end{equation}
Here, \(\Sigma_{\rm SYK}\) is the usual on-site self-energy, while \(\Sigma_{\rm hop}\) encodes the spatial deformation. At the level of chaos, for the bilinear hopping deformation considered here, the corresponding retarded ladder kernel factorizes into an on-site part and a hopping-induced part,
\begin{equation}\label{eq:KRfactorized}
K_R(k,\omega)=K_{\rm SYK}(\omega)+K_{\rm hop}(k,\omega),
\qquad
K_{\rm hop}(k,\omega)=\Gamma(k)\,\Pi(\omega),
\end{equation}
where \(K_{\rm SYK}(\omega)\) is the effective single-site retarded kernel in the IR regime, \(\Pi(\omega)\) is the usual RA fermion bubble built from the on-site real-time two-point functions,
\begin{equation}\label{eq:Pidef}
\Pi(\omega)=\int \frac{d\Omega}{2\pi}
\left[
G^{R}(\Omega+\omega)G^{<}(\Omega)+
G^{<}(\Omega+\omega)G^{A}(\Omega)
\right],
\end{equation}
and \(\Gamma(k)\) is the momentum-space structure factor associated with the hopping. Thus, all spatial information enters through \(\Gamma(k)\), while the purely temporal chaotic dynamics is inherited from the on-site SYK sector. As usual, the zero-momentum contribution may be absorbed into the effective on-site kernel, so that the momentum dependence relevant for chaotic spreading is controlled by the nontrivial \(k\)-dependence of \(\Gamma(k)\).

The resummed expression \eqref{eq:ladder_resummed} shows that the OTOC develops an exponentially growing mode whenever the denominator vanishes. We therefore parametrize the pole as
\begin{equation}
\omega=i\Lambda(k),
\end{equation}
with \(\Lambda(0)=\lambda_L\), and determine it from
\begin{equation}\label{eq:polecondition}
1-K_R\!\left(k,i\Lambda(k)\right)=0.
\end{equation}
Expanding at small momentum then gives
\begin{equation}\label{eq:Lambda_general_expand}
\Lambda(k)=\lambda_L+\Lambda_1 k+\Lambda_2 k^2+\cdots.
\end{equation}
For a standard nearest-neighbor hopping term, the linear term vanishes and one recovers the usual diffusive form \(\Lambda(k)=\lambda_L-Dk^2/2+\cdots\), whereas more general deformations can generate non-analytic or even complex momentum dependence. Passing back to real space, the connected OTOC behaves as
\begin{equation}\label{eq:OTOCrealspace}
F(t,x)=1-\frac{1}{N}\int \frac{dk}{2\pi}\,
e^{ikx}e^{\Lambda(k)t},
\end{equation}
so the long-distance spread of chaos is entirely controlled by the small-\(k\) structure of \(\Lambda(k)\).

We now specialize to the long-range deformation \eqref{eq:Hhop}. For this choice, the hopping self-energy becomes
\begin{equation}
\Sigma_{\rm hop}(k,\tau)=J_1^2S_\alpha(k)G(\tau),
\quad
S_\alpha(k)\equiv\sum_{r\neq 0}\frac{e^{ikr}}{|r|^{2\alpha}},
\label{eq:selfenergy_def}
\end{equation}
so in this case the structure factor appearing in \eqref{eq:KRfactorized} is simply
\begin{equation}
\Gamma(k)=S_\alpha(k).
\end{equation}
The power \(2\alpha\) appears because \(\Sigma_{\rm hop}\) is generated by a disorder average over two hopping insertions with variance \(\langle (I^{(a)}_{jk})^2\rangle=J_1^2\), which yields a kernel proportional to \(|j-k|^{-2\alpha}\). The small-\(k\) behavior of \(S_\alpha(k)\) is determined by the discrete Fourier transform of this power-law profile. Explicitly, on an infinite line
\(
S_\alpha(k)=2\sum_{r=1}^\infty \frac{\cos(kr)}{r^{2\alpha}},
\)
and for \(1<\alpha<2\), one finds the non-analytic expansion
\begin{equation}
S_\alpha(k)=S_\alpha(0)-c_\alpha |k|^{\alpha-1}+\mathcal{O}(k^2),\quad c_\alpha>0.
\label{eq:Salpha_smallk}
\end{equation}
Substituting this into the pole condition \eqref{eq:polecondition} gives, to leading order in small \(k\),
\begin{equation}\label{eq:LambdaDisp}
\Lambda(k)=\lambda_{L}-\Gamma_{\alpha}|k|^{\alpha-1},
\quad
\Gamma_{\alpha} = c_{\alpha}\lambda_{L}\left(\frac{J_{1}}{J_{0}}\right)^{2}.
\end{equation}
Equation \eqref{eq:LambdaDisp} is the long-range dispersion relation for the chaotic growth. At \(k=0\), \(\Lambda(0)=\lambda_L\), while for \(k\neq0\), the exponent is reduced by a non-analytic term. Compared with the familiar diffusive \(k^2\) correction of short-range chains, the softer \(|k|^{\alpha-1}\) dependence implies that modes with finite momentum continue to grow almost as fast as the zero-momentum one, which in turn accelerates the spatial spread of chaos.

To make this more concrete, let us consider the spatiotemporal behavior of the OTOC. Using \eqref{eq:OTOCrealspace} together with \eqref{eq:LambdaDisp}, and setting \(x=vt\), the exponent in the integrand becomes \(\Phi(k)=ikvt+\Lambda(k)t\). For large \(t\), the integral can be evaluated by a saddle-point approximation. The dominant contribution comes from \(k=k_*\) satisfying
\begin{equation}
iv+\Lambda'(k_*)=0.
\end{equation}
Solving this equation and substituting back into \(\Phi(k)\), one finds
\begin{equation}
\mathrm{Re}\Phi(k_*)=
\left( \lambda_L-A_\alpha\Gamma_\alpha^{-\tfrac{1}{\alpha-1}}
      v^{\tfrac{\alpha-1}{\alpha-2}}\right) t,
\end{equation}
with
\begin{equation}
A_\alpha=\left|\frac{\sin\frac{\pi}{2-\alpha}+(\alpha-1)\cos\frac{\pi(\alpha-1)}{2(2-\alpha)}}{(\alpha-1)^{\frac{\alpha-1}{\alpha-2}}}\right|.
\end{equation}
The butterfly velocity is then determined by the condition \(\mathrm{Re}\Phi(k_*)=0\), which yields
\begin{equation}\label{vBresultChain}
  v_B(\alpha)=
  \Gamma_\alpha^{\tfrac{1}{\alpha-1}}
  \left(\frac{\lambda_L}{A_\alpha}\right)^{\tfrac{\alpha-2}{\alpha-1}},
\quad 1<\alpha.
\end{equation}
This reproduces the result quoted earlier. We see that \(v_B(\alpha)\) increases as \(\alpha\) decreases; in particular, as the power-law tail thickens (\(\alpha\to1^+\)), the butterfly velocity grows without bound,
\begin{equation}
v_B(\alpha\to 1^+)\sim(\alpha-1)\left(\frac{J_1}{J_0}\right)^{\tfrac{2}{\alpha-1}}\to\infty.
\end{equation}
Intuitively, when interactions become nearly all-to-all, even distant parts of the system can scramble each other almost immediately, so information can propagate arbitrarily fast.

It is worth commenting on the case $\alpha\leq 1$. In this regime, the momentum-dependent correction in \eqref{eq:LambdaDisp} would naively diverge as $k\to0$. However, a careful analysis shows that the fractional change due to the momentum-dependent part vanishes in the IR: $[S_{\alpha}(k)-S_{\alpha}(0)]/S_{\alpha}(0)\to0$ as $k\to0$. In other words, for $\alpha\le1$, the kernel becomes effectively momentum-independent in the long-wavelength limit. The saddle-point condition then collapses to $k_*=0$, so $\mathrm{Re}\Phi=\lambda_L t$ independently of $v$. There is therefore no finite velocity at which the real part of the exponent vanishes, and hence no finite butterfly velocity. Physically, when $\alpha\le1$, even the average hopping range diverges, meaning that the chain becomes effectively all-to-all in a single microscopic time step. The notion of a sharp butterfly front, and hence any finite Lieb--Robinson-type velocity, ceases to exist in this extreme long-range regime.

Here, we summarize the scaling of $v_{B}$ in four distinct regimes:
\begin{itemize}
\item \textbf{Diffusive regime} ($\alpha>2$): the kernel acquires the familiar quadratic correction $\Lambda(k)\approx\lambda_{L}-Dk^{2}$. Operator growth is a standard random walk and a sharp butterfly cone with finite $v_{B}=\sqrt{D\lambda_{L}}$ emerges.
\item \textbf{Marginal regime} ($\alpha=2$): the dispersion softens to linear form $\Lambda(k)\approx\lambda_{L}-\gamma|k|$, so the chaotic front broadens only logarithmically and $v_{B}$ picks up a slow $\log t$ enhancement.
\item \textbf{Super-diffusive regime} ($1<\alpha<2$): the non-analytic term $\propto|k|^{\alpha-1}$ signals L\'evy flights \cite{shlesinger1995levy}. Rare long hops propel the chaos front ahead of any local light cone, with $v_{B}$ following the power law of Eq.~\eqref{vBresultChain} and diverging as $\alpha\to1^{+}$.
\item \textbf{Instant-scrambling regime} ($\alpha\le1$): the $k$-dependence of the kernel disappears; each site talks to a macroscopic fraction of the chain in one step. The concept of a finite butterfly velocity breaks down.
\end{itemize}

Our result realizes one of the qualitative traits found for dS black holes, namely, the superluminal behavior as the bulk black hole radius shrinks. In this regime, the dual boundary interactions must become increasingly non-local in space so that the resulting $v_{B,\mathrm{bh}}$ can exceed the speed of light. We stress that the Lyapunov exponent remains maximal for all values of $\alpha$; it is the spatial spread, not the local scrambling rate, that encodes the transition from diffusive to superluminal behavior.

\subsubsection{Non-Hermitian DSSYK-type chain and the imaginary butterfly velocity}

Inspired by the previous results, we now propose a toy model that can capture the purely \emph{imaginary} butterfly velocity associated with the cosmological horizon, $v_{B,\mathrm{c}}$. In our bulk analysis, the cosmological horizon in dS was found to exhibit an imaginary $v_B$ (see Section~\ref{sec43}), indicating the absence of a well-defined causal front for the chaotic spreading of information. Such behavior is not entirely unexpected: several early proposals for dS holography have invoked non-unitary dynamics together with unusual Hermiticity conditions~\cite{Strominger:2001pn,Witten:2001kn,Bousso:2001mw,Balasubramanian:2002zh,Anninos:2011ui}. This motivates us to consider a non-Hermitian deformation of the DSSYK chain, which may be viewed as the effective dynamics of an open quantum system after tracing out the environment. As in the previous subsection, we restrict attention to the IR conformal regime.

Our starting point is again a chain of $L$ identical sites labeled $j=1,\dots,L$. However, to allow for non-Hermitian hopping, we work with a slightly more general version of the chain in which the fields $\psi_{j,a}$ are complex fermions. Each site hosts $N$ such fermions, with an on-site Hamiltonian given by the natural complex-fermion generalization of SYK$_q$,
\begin{equation}\label{eq:cSYKq}
H^{(j)}_{\text{SYK}}
=
\sum_{\substack{1\le a_1<\cdots<a_{q/2}\le N\\
1\le b_1<\cdots<b_{q/2}\le N}}
J^{(j)}_{a_1\cdots a_{q/2};\,b_1\cdots b_{q/2}}\,
\psi^\dagger_{j,a_1}\cdots \psi^\dagger_{j,a_{q/2}}
\psi_{j,b_1}\cdots \psi_{j,b_{q/2}},
\end{equation}
where $q$ is taken to be even, and the random couplings satisfy
\begin{equation}
J^{(j)}_{a_1\cdots a_{q/2};\,b_1\cdots b_{q/2}}
=
\Bigl(
J^{(j)}_{b_1\cdots b_{q/2};\,a_1\cdots a_{q/2}}
\Bigr)^*,\qquad
\Bigl\langle
\bigl|J^{(j)}_{a_1\cdots a_{q/2};\,b_1\cdots b_{q/2}}\bigr|^2
\Bigr\rangle
=
\frac{\left(\frac q2\right)!^2 J_0^2}{N^{q-1}}.
\end{equation}
This complex-fermion SYK$_q$ model lies in the same IR chaotic universality class as the standard SYK$_q$~\cite{Bulycheva:2017uqj}; in particular, it saturates the chaos bound for $\lambda_L$. We then add a nearest-neighbor hopping term that is generically non-Hermitian, and later distinguish between $PT$-symmetric and $PT$-breaking choices. Concretely, we introduce
\begin{equation}
  H_{\text{hop}}
    =
    \sum_{j=1}^{L}\sum_{a=1}^{N}
      \Bigl(
        \mu^{(R)}_{j,a}\,\psi^\dagger_{j+1,a}\psi_{j,a}
        +\mu^{(L)}_{j,a}\,\psi^\dagger_{j,a}\psi_{j+1,a}
      \Bigr),
\end{equation}
where $\mu^{(R)}_{j,a}$ and $\mu^{(L)}_{j,a}$ are independent random amplitudes for right- and left-moving hopping, respectively. We choose these couplings from different complex distributions such that, in the $PT$-broken case, $\mu^{(L)}_{j,a}\neq (\mu^{(R)}_{j,a})^*$. For definiteness, let them be Gaussian distributed with zero mean and variances
\begin{equation}
  \bigl\langle \mu^{(R)}_{j,a}(\mu^{(R)}_{j',a'})^*\bigr\rangle
    =
    \mu_{R}^{2}\,\delta_{jj'}\delta_{aa'},
  \quad
  \bigl\langle \mu^{(L)}_{j,a}(\mu^{(L)}_{j',a'})^*\bigr\rangle
    =
    \mu_{L}^{2}\,\delta_{jj'}\delta_{aa'},
\end{equation}
where $\mu_{R,L}$ control the right- and left-hopping strengths. At large $N$, the local on-site dynamics continues to control the temporal growth, so the maximal Lyapunov exponent remains $\lambda_L = 2\pi/\beta$. The role of the hopping terms is instead to endow the OTOC with a non-trivial spatial dispersion $\Lambda(k)$.

Proceeding exactly as in the previous subsection, we analyze the connected OTOC by summing the ladder diagrams built from the dressed two-point functions. For the bilinear hopping deformation considered here, the retarded ladder kernel again takes the form
\begin{equation}
  K_{R}(k,\omega)
  =
  K_{\text{SYK}}(\omega)+K_{\text{hop}}(k,\omega),
  \qquad
  K_{\text{hop}}(k,\omega)=\Gamma(k)\Pi(\omega),
\end{equation}
where \(K_{\text{SYK}}(\omega)\) is the effective single-site retarded kernel in the IR regime and \(\Pi(\omega)\) is the RA fermion bubble defined in Eq.~\eqref{eq:Pidef}. To leading order in \(1/N\), the on-site two-point functions remain those of a single site, so the non-Hermitian character of the model enters entirely through the momentum-space structure factor
\begin{equation}\label{def:Gamma}
  \Gamma(k)=\mu_{R}^{2}e^{ik}+\mu_{L}^{2}e^{-ik}.
\end{equation}
If the hopping is $PT$-symmetric, a simple choice is to take equal left- and right-hopping statistics, $\mu_L=\mu_R=\mu$, in which case \(\Gamma(k)=2\mu^2\cos k\) is an even real function of \(k\). In contrast, in the $PT$-broken case, \(\Gamma(k)\) is generally complex and satisfies \(\Gamma(-k)\neq \Gamma(k)\). This asymmetry between \(+k\) and \(-k\) is the source of the novel chaotic propagation described below. As in the previous subsection, the zero-momentum contribution may be absorbed into the effective on-site kernel, so that the momentum dependence relevant for chaotic spreading is controlled by the nontrivial \(k\)-dependence of \(\Gamma(k)\).

The OTOC amplitude therefore obeys the same resummed ladder expression as before,
\begin{equation}
f(k,\omega)=\frac{K_R(k,\omega)}{1-K_R(k,\omega)}\,f^{(0)}(k,\omega),
\end{equation}
and the growth exponent \(\Lambda(k)\) is again determined by the pole condition
\begin{equation}
1-K_R\!\left(k,i\Lambda(k)\right)=0.
\end{equation}
Expanding around \(k=0\), we write
\begin{equation}
  \Lambda(k)=\lambda_{L}+\Delta\Lambda(k),\qquad \Delta\Lambda(0)=0,
\end{equation}
so that, to leading order,
\begin{equation}
  \Delta\Lambda(k)
  \approx-\frac{K_{\text{hop}}\!\bigl(k,i\lambda_{L}\bigr)}
                {i K_{\text{SYK}}'(i\lambda_{L})}.
\end{equation}
Thus the sign and nature of \(\Delta\Lambda(k)\) are controlled directly by the structure factor \(\Gamma(k)\).

We can now distinguish two cases.

\paragraph{$PT$-symmetric hopping:} If $\mu_{R}=\mu_{L}=\mu$, then
\begin{equation}
\Gamma(k) = 2\mu^2 \cos k,
\end{equation}
which is real and even. In this case, \(K_{\text{hop}}(k,i\lambda_L)\) is an even function of \(k\) with a purely real expansion. Consequently, the linear term vanishes and the leading correction is quadratic:
\begin{equation}
  \Lambda(k)=\lambda_{L}-\frac{\mathcal{D}}{2}k^{2}+O(k^{4}),
  \qquad
  v_{B}=\sqrt{2\lambda_L \mathcal{D}}.
\end{equation}
Here \(\mathcal{D}\) is a positive constant, proportional to \(\mu^2\Pi(i\lambda_L)/|K'_{\text{SYK}}(i\lambda_L)|\), which plays the role of a diffusion constant for chaos. The butterfly velocity is therefore real, in agreement with the standard expectation for a chaotic system with a ballistically propagating front accompanied by diffusive broadening.

\paragraph{$PT$-breaking hopping:} When \(\mu_R\neq\mu_L\), one instead has
\begin{equation}
  \Gamma(k)=\mu_R^2 e^{ik}+\mu_L^2 e^{-ik},
\end{equation}
whose small-\(k\) expansion reads
\[
\Gamma(k)\approx \mu_R^2+\mu_L^2+i(\mu_R^2-\mu_L^2)k-\frac{1}{2}(\mu_R^2+\mu_L^2)k^2+\cdots.
\]
In this case, \(K_{\text{hop}}(k,i\lambda_L)\) generically develops an imaginary part linear in \(k\), and \(\Delta\Lambda(k)\) therefore acquires a linear imaginary contribution. The growth exponent therefore takes the form
\begin{equation}
  \Lambda(k)=
    \lambda_{L}
    +i v_{B}k
    -\frac{\mathcal{D}}{2}k^{2}
    +O(k^{3}),
  \qquad
  v_{B}=i\,\frac{\left(\mu_R^2-\mu_L^2\right)\Pi(i\lambda_L)}{|K'_{\text{SYK}}(i\lambda_L)|}.
\end{equation}
We thus find that the butterfly ``velocity'' in the $PT$-broken chain is \emph{purely imaginary}. Equivalently, while the real part of \(\Lambda(k)\) remains maximal to leading order, the asymmetric hopping induces an imaginary phase proportional to \(k\). In other words, the leading effect of the non-Hermitian deformation is not to suppress the temporal growth rate, but to modify its spatial structure through an oscillatory phase.

The butterfly velocity therefore becomes purely imaginary in the $PT$-broken DSSYK-type chain, closely paralleling the behavior inferred for the cosmological horizon, \(v_{B,\text{c}}\), in de Sitter space. Operationally, an imaginary \(v_B\) means that there is no well-defined finite speed delimiting a causal light cone for operator growth. Instead of a sharp front, the spatial profile is governed by oscillatory or exponentially attenuated behavior away from the source. Nevertheless, the Lyapunov exponent remains maximal, \(\lambda_L=2\pi/\beta\). Thus, while the temporal rate of exponential growth is unchanged by the non-Hermitian deformation, the spatial pattern of that growth is profoundly altered by the asymmetric hopping. The $PT$-broken DSSYK-type chain therefore provides a useful toy example of a system with maximal Lyapunov exponent but no sensible finite butterfly velocity, closely mimicking the cosmological-horizon scenario in de Sitter space.

%
\section{Conclusions}\label{sec5}

In this paper, we studied the pole-skipping phenomenon in cosmological spacetimes, focusing in particular on asymptotically de Sitter black hole geometries of the Schwarzschild--de Sitter (SdS) type. We identified the leading pole-skipping points for probe scalar and Maxwell fields, as well as for linearized gravitational perturbations, both at the black hole event horizon and at the de Sitter cosmological horizon. As in the more familiar AdS/CFT setting, the gravitational sound mode exhibits a distinguished `chaotic' pole-skipping point in the upper half of the complex $\omega$-plane. Remarkably, this phenomenon is present at both horizons. Interpreting these special points in terms of the Lyapunov exponent and butterfly velocity $v_B$ of a hypothetical dual many-body system, we found unconventional regimes in which $v_B$ becomes either superluminal or purely imaginary.

To corroborate these results and further clarify their implications for the spread of quantum chaos in a hypothetical dual description of de Sitter space, we also carried out a gravitational shock wave analysis, which in standard holography is related to out-of-time-ordered correlators (OTOCs). Extracting the Lyapunov exponents and butterfly velocities from the shock wave solutions, we found precise agreement with the pole-skipping analysis at both horizons of the SdS geometry. This provides additional evidence that pole-skipping is a reliable diagnostic of high-energy scattering near horizons, and further supports its interpretation as a probe of the chaotic sector of a putative holographic dual.

Although the construction of an explicit holographic dual for de Sitter spacetimes remains an open challenge, our results reveal a number of intriguing parallels and important differences between the chaotic properties of AdS and dS horizons. Motivated by these findings, we proposed a tentative microscopic framework that could account for the superluminal and imaginary butterfly velocities encountered in our analysis. In particular, we discussed double-scaled Sachdev--Ye--Kitaev (DSSYK) chains with long-range interactions and non-Hermitian deformations as simple toy models capable of reproducing some of the qualitative features suggested by the bulk calculations.

It is worth emphasizing, however, that our use of DSSYK is intentionally modest. We do not claim that the long-range or non-Hermitian DSSYK chains considered here provide a complete microscopic dual of de Sitter space. Rather, they should be viewed as toy models that reproduce qualitative features of the chaotic sector: the double-scaled on-site dynamics preserves the standard maximal Lyapunov exponent, while the deformations modify the momentum dependence of the growth exponent and hence the spatial spread of chaos, allowing for superluminal or purely imaginary butterfly velocities. In this sense, our proposal is complementary to earlier de Sitter-inspired SYK constructions. The distinctive feature of our setup is a standard maximal $\lambda_L$ together with unconventional spatial propagation encoded in $v_B$. This also differentiates our construction from centaur-based models, which involve flow geometries with AdS asymptotics and a de Sitter interior and therefore need not reproduce the same chaotic observables quantitatively.

More broadly, our results suggest that any future microscopic description of de Sitter horizon dynamics may have to incorporate ingredients that are uncommon from the perspective of ordinary local unitary quantum field theory, such as sufficiently nonlocal interactions or some controlled form of non-Hermiticity. We hope that further studies of SYK-like models, Brownian or non-Hermitian variants, and other higher-dimensional nonlocal systems will help sharpen these ideas and ultimately contribute to a better understanding of holography in de Sitter space and, more generally, in cosmological settings.

%
\acknowledgments
SG is supported by the STFC Ernest Rutherford Fellowship ST/T00388X/1. His work is also supported by the research programme P1-0402 and the project J7-60121 of Slovenian Research Agency (ARIS).
HSJ and JFP are supported by the Spanish MINECO ‘Centro de Excelencia Severo Ochoa' program under grant SEV-2012-0249, the Comunidad de Madrid ‘Atracci\'on de Talento’ program (ATCAM) grant 2020-T1/TIC-20495, the Spanish Research Agency via grants CEX2020-001007-S and PID2021-123017NB-I00, funded by MCIN/AEI/10.13039/501100011033, and ERDF A way of making Europe.
HSJ was supported by an appointment to the JRG Program at the APCTP through the Science and Technology Promotion Fund and Lottery Fund of the Korean Government. HSJ was also supported by the Korean Local Governments -- Gyeongsangbuk-do Province and Pohang City.
All authors contributed equally to this paper and should be considered as co-first authors.

%
\appendix
\section{Analytic correlators in empty de Sitter space}
In this appendix, we demonstrate that the pole-skipping point identified through the near-horizon analysis in the main text is consistent with the corresponding point derived from the singular behavior of correlators in de Sitter space. For this purpose, we focus on the case in which the bulk differential equation admits an exact analytic solution.

In the case of empty de Sitter space, corresponding to \eqref{Sol:f} with vanishing mass parameter $M=0$, the Klein-Gordon equation~\eqref{Eq:scalar} allows for an analytic solution~\cite{Abdalla:2002hg,Lopez-Ortega:2006aal}. For completeness and consistency with our notation, we rederive this solution explicitly in the following.

When the mass parameter is set to zero, the Klein-Gordon equation \eqref{Eq:scalar} reduces to a hypergeometric differential equation of the form
\begin{equation}
\label{eq:hypergeometric}
z(z-1)F''(z)+\left[(1+A+B)z-C\right] F'(z)+A B F(z),
\end{equation}
where $z=r^2/L^2$, $F$ arises from the definition
\begin{equation}
\bar{\phi}(z)=e^{-i\omega r_*(z)}z^{\alpha} (1-z)^\beta F(z),
\end{equation}
and the coefficients $(\alpha, \beta, A, B, C)$ satisfy
\begin{equation}
\begin{split}
\label{eq:coefshyp}
4 \alpha ^2+2 \alpha (d-2)-k_S^2 &= 0,\\
4 \beta ^2+L^2 \omega ^2 &= 0,\\
A+B&=1+i \omega L+\sqrt{(d-2)^2+4 k_S^2},\\
AB&=\frac{1}{4} \left(1+i\omega L+\sqrt{(d-2)^2+4 k_S^2}\right)^2-\left[\left(\frac{d}{4}\right)^2-\left(\frac{m_\phi L}{2}\right)^2\right],\\
C&=1+\sqrt{(d-2)^2+4 k_S^2}.
\end{split}
\end{equation}
Note that any combination of $\alpha$ and $\beta$ satisfying Eqs.~\eqref{eq:coefshyp} is equivalent. For convenience, we choose 
\begin{equation}
\alpha=-\frac{1}{4}\left((d-2)+\sqrt{(d-2)^2+4k_S^2}\right) ,\quad
\beta=-i\frac{\omega L}{2}.
\end{equation}
The scalar field solution to \eqref{eq:hypergeometric} is then a linear combination of hypergeometric functions as
\begin{equation}
\label{sol:F}
F(z)={}_2F_1(A,B;C;z) + \mathcal{A}  z^{1-C}{}_2F_1(1+A-C,1+B-C;2-C;z),
\end{equation}
where the constant $\mathcal{A}$ can be chosen to satisfy ingoing boundary condition at $z=1$. Next, we expand \eqref{sol:F} near $z=0$, the worldlines of inertial observers at the spatial poles (or the location of a stationary observer at the center of the universe), as
\begin{equation}
\bar{\phi}(z)\approx \bar{\phi}^{(L)}z^{\alpha}+\ldots +\bar{\phi}^{(SL)}z^{-\frac{d-2}{2}-\alpha}+,
\end{equation}
where $\bar{\phi}^{(L)}$ and $\bar{\phi}^{(SL)}$ are the leading and subleading coefficients, respectively.
Using this asymptotic behavior of the hypergeometric function, one may obtain the correlator or the retarded Green's function of a dual scalar operator as
\begin{equation}
G^R_{\mathcal{O}\mathcal{O}} = \frac{\bar{\phi}^{(SL)}}{\bar{\phi}^{(L)}}\propto \frac{\Gamma(1+A-C)\Gamma(1+B-C)}{\Gamma(A)\Gamma(B)},
\end{equation}
where $\Gamma(x)$ is the Gamma function, which diverges at non-positive integers. Consequently, pole-skipping occurs when both the numerator and denominator vanish simultaneously, i.e., $G^R_{\mathcal{O}\mathcal{O}} \sim$ `$0/0$'. The leading pole-skipping point can then be found at  
\begin{equation}
\omega=-\frac{i}{L}, \quad k_S^2=d-1-m_\phi^2L^2,
\end{equation}
which agrees with the result from Eq.~\eqref{psp:scalarcosh}, where $r_\mathrm{c}=L$, $T_\mathrm{c} = 1/2\pi L$, and $f'(r_\mathrm{c})=-2/L$.

%
\bibliographystyle{JHEP}

\providecommand{\href}[2]{#2}\begingroup\raggedright\endgroup

\end{document}